\documentclass{llncs}
\usepackage{epstopdf}
\usepackage{url}
\usepackage{multicol}
\usepackage{epsfig}
\usepackage{subfigure}
\newtheorem{defn}{Definition}
\usepackage{graphicx}
\usepackage{algorithm}
\usepackage[noend]{algorithmic}
\usepackage{amsmath}
\usepackage{colortbl}
\usepackage{booktabs}
\usepackage{amssymb}
\usepackage{mathrsfs}

\usepackage{makeidx}  % allows for indexgeneration
\floatname{algorithm}{Algorithm}

\begin{document}

%%\frontmatter   
%
%\title{Flip the Cloud: Cyber-Physical Signaling Games in the Presence of Advanced Persistent Threats}
%%\subtitle{[Extended Abstract]
%%\titlenote{A full version of this paper is available as
%%\textit{Author's Guide to Preparing ACM SIG Proceedings Using
%%\LaTeX$2_\epsilon$\ and BibTeX} at
%%\texttt{www.acm.org/eaddress.htm}}}
%%

\frontmatter   
\title{Flip the Cloud: Cyber-physical Signaling Games in the Presence of Advanced Persistent Threats}
\author{Jeffrey Pawlick, Sadegh Farhang, and Quanyan Zhu}
\institute{Department of Electrical and Computer Engineering, \\Polytechnic School of Engineering, New York University, New York, USA\
\email{\{jpawlick,farhang,quanyan.zhu\}@nyu.edu}\\}

\maketitle              

\begin{abstract}
Access to the cloud has the potential to provide scalable and cost effective enhancements of physical devices through the use of advanced computational processes run on apparently limitless cyber infrastructure. On the other hand, cyber-physical systems and cloud-controlled devices are subject to numerous design challenges; among them is that of security. In particular, recent advances in adversary technology pose Advanced Persistent Threats (APTs) which may stealthily and completely compromise a cyber system. In this paper, we design a framework for the security of cloud-based systems that specifies when a device should trust commands from the cloud which may be compromised. 
This interaction can be considered as a game between three players: a cloud defender/administrator, an attacker, and a device. We use traditional signaling games to model the interaction between the cloud and the device, and we use the recently proposed \texttt{FlipIt} game to model the struggle between the defender and attacker for control of the cloud. Because attacks upon the cloud can occur without knowledge of the defender, we assume that strategies in both games are picked according to prior commitment.  
%% The following text is an edit after the initial submission:
%We consider a three-player model of cloud-controlled devices.
%The bottom layer represents a physical device which can be described
%by a dynamic system.  The top layer consists of a cyber resource available
%to the device.  This layer is managed by an administrator 
%and subject to APT from an attacker. We model the conflict over control
%of the top layer
%using the recently proposed \texttt{FlipIt} game.  Finally, the 
%interaction layer between the top and bottom layers can be modeled
%by a traditional signaling game. Because the attacker and defender can 
%capture the cloud without knowledge of the other two players, we assume 
%that strategies in both the signaling and \texttt{FlipIt} games are picked
%according to prior commitment. %% End of new text
This framework requires a new equilibrium concept, which we call \emph{Gestalt Equilibrium}, a fixed-point that expresses the interdependence of the signaling and \texttt{FlipIt} games.  We present the solution to this fixed-point problem under certain parameter cases, and illustrate an example application of cloud control of an unmanned vehicle.  Our results contribute to the growing understanding of cloud-controlled systems.

\end{abstract}
%%%%%%%%%%%%%%%%%%%%%%%%%%%%%%%%%%%%%%%%%%%%%%%%%%%%%%%%%%
\section{Introduction}
\label{sec:Intro}
Advances in computation and information analysis have expanded the capabilities
of the physical plants and devices in cyber-physical systems (CPS)\cite{baheti2011cyber,lee2008cyber}.
Fostered by advances in cloud computing, CPS have garnered significant attention from both industry and academia.  Access to the cloud gives
administrators the opportunity to build virtual machines that provide to 
computational resources with precision, scalability, and accessibility. 

Despite the advantages that cloud computing provides, it also has
some drawbacks. They include - but are not limited to - accountability,
virtualization, and security and privacy concerns. In this paper,
we focus especially on providing accurate signals to a cloud-connected
device and deciding whether to accept those signals in the face of
security challenges.

Recently, system designers face security challenges in the form of
\emph{Advanced Persistent Threats} (\emph{APTs}) \cite{tankard2011advanced}.
APTs arise from sophisticated attackers who can infer a user's cryptographic
key or leverage zero-day vulnerabilities in order to completely compromise
a system without detection by the system administrator \cite{portokalidis2006argos}.
This type of stealthy and complete compromise has demanded new types
of models \cite{bowers2012defending,vanDijk2013Flip} for prediction
and design.

In this paper, we propose a model in which a device decides whether to trust
commands from a cloud which is vulnerable to APTs and may fall under 
adversarial control.  We synthesize a mathematical
framework that enables devices controlled by the cloud to intelligently
decide whether to obey commands from the possibly-compromised cloud
or to rely on their own lower-level control. 

%%% Edits after inital submission to the following:
We model the cyber layer of the cloud-based system
using the recently proposed \texttt{FlipIt} game 
\cite{bowers2012defending,vanDijk2013Flip}. 
This game is especially suited for studying systems under APTs.
We model the interaction between the cloud and the connected device
using a signaling game, which provides a framework for modeling dynamic
interactions in which one player operates based on a belief about the 
private information of the other. A significant body of
research has utilized this framework for security \cite{carroll2011game,farhang2014dynamic,pawlick2015deception,zhuang2010modeling,casey2014}.  The signaling and \texttt{FlipIt} games are coupled,
because the outcome of the \texttt{FlipIt} game determines the likelihood
of benign and malicious attackers in the robotic signaling game. Because the attacker is able to compromise
the cloud without detection by the defender, we consider the strategies
of the attacker and defender to be chosen with \emph{prior commitment}.
 The circular dependence in our game requires a new equilibrium
concept which we call a \emph{Gestalt equilibrium}%
\footnote{Gestalt is a noun which means something that is composed of multiple
parts and yet is different from the combination of the parts \cite{gestalt}.%
}.  We specify the parameter cases under which the Gestalt equilibrium
varies, and solve a case study of the game to give an idea of how
the Gestalt equilibrium can be found in general. Our proposed framework has versatile applications to different cloud-connected systems such as urban traffic control, drone delivery, design of smart
homes, etc. We study one particular application in this paper:ef control of an
unmanned vehicle under the threat of a compromised cloud.

Our contributions are summarized as follows:
\begin{description}
\item[i)]We model the interaction of the attacker, defender/cloud administrator, and cloud-connected device by introducing a novel game consisting of two
coupled games: a traditional signaling game and the recently proposed
\texttt{FlipIt} game. 
\item[ii)]We provide a general framework by which a device
connected to a cloud can decide whether to follow its own limited
control ability or to trust the signal of a possibly-malicious cloud.
\item[iii)]We propose a new equilibrium definition for this combined
game: Gestalt equilibrium, which involves a fixed-point in the mappings
between the two component games.
\item[iv)]Finally, we apply our framework to the problem of unmanned
vehicle control.
\end{description}

In the sections that follow, we first outline the system model, then
describe the equilibrium concept. Next, we use this concept to find
the equilibria of the game under selected parameter regimes. Finally,
we apply our results to the control of an unmanned vehicle. In each
of these sections, we first consider the signaling game, then consider
the \texttt{FlipIt} game, and last discuss the synthesis of the two
games. Finally, we conclude the paper and suggest areas for future
research.
%%%%%%%%%%%%%%%%%%%%%%%%%%%%%%%%%%%%%%  
\section{System Model}
\label{sec:SysMod}
We model a cloud-based system in which a cloud is subject to
APTs. In this model, an \emph{attacker}, denoted by $\mathcal{A}$,
capable of APTs can pay an attack cost to completely compromise
the cloud without knowledge of the cloud defender. The \emph{defender},
or cloud administrator, denoted by $\mathcal{D}$, does not observe these attacks, but has
the capability to pay a cost to reclaim control of the cloud. The
cloud transmits a message to a \emph{robot} or other device, denoted 
by $\mathcal{R}$. The device may follow this command, but it is also
equipped with an on-board control system for autonomous operation.
It may elect to use its autonomous operation system rather than
obey commands from the cloud.

\begin{figure}[t]
\begin{center}
\includegraphics[width=0.55\columnwidth]{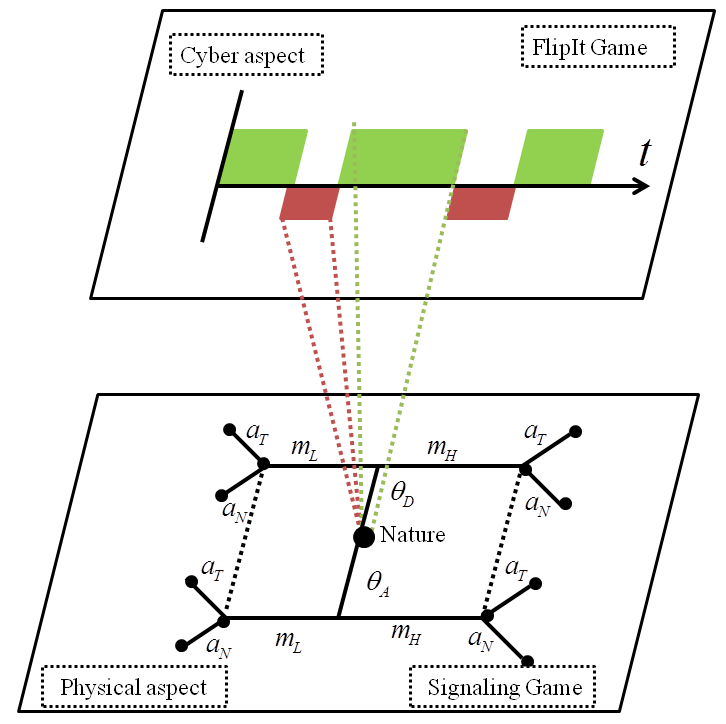}

\caption{The \texttt{CloudControl} game. The \texttt{FlipIt} game models the interaction between an attacker and a cloud administrator for control of the cloud. The outcome of this game determines the type of the cloud in a signaling game in which the cloud conveys commands to the robot or device.  The device then decides whether to accept these commands or rely on its own lower-level control. The \texttt{FlipIt} and signaling games are played concurrently.}
\label{fig:SysModel}
\end{center}
\end{figure}

This scenario involves two games: the \texttt{FlipIt}
game introduced in \cite{vanDijk2013Flip},
and the well-known signaling game. The \texttt{FlipIt} game takes
place between the attacker and cloud defender, while the signaling
game takes place between the possibly-compromized cloud and the device. For brevity, denote the \texttt{FlipIt} game by $\mathbf{G_{F}}$,
the signaling game by $\mathbf{G_{S}}$, and the combined game - call
it \texttt{CloudControl} - by $\mathbf{G_{CC}}$ as shown in Fig. \ref{fig:SysModel}. 
%Further, let $\mathcal{A}$, $\mathcal{D}$, and $\mathcal{R}$ denote the attacker, cloud defender/administrator and robotic resource, respectively. 
In the next subsections, we formalize
this game model.

\subsection{Cloud-Device Signaling Game }

%In signaling games, a \emph{type} is realized from a random variable
%according to some known probability distribution. A \emph{sender}
%observes the type, and sends a \emph{message} to a receiver. The receiver
%observes the message, but not the type, and chooses an \emph{action}.
%The sender and receiver obtain payoffs that are determined by a function
%of the type, message, and action.

%\subsubsection{Types, Messages, Actions}

Let $\theta$ denote the type of the cloud. Denote \emph{compromized}
and \emph{safe} types of clouds by $\theta_{\mathcal{A}}$ and $\theta_{\mathcal{D}}$
in the set $\Theta$. Denote the probabilities that $\theta=\theta_{\mathcal{A}}$
and that $\theta=\theta_{\mathcal{D}}$ by $p$ and $1-p$. Signaling
games typically give these probabilities \emph{apriori}, but in \texttt{CloudControl}
they are determined by the equilibrium of the \texttt{FlipIt} game
$\mathbf{G_{F}}$. 

%In general, the space of the sender's commands could be large or even continuous. 
%We consider that the robot implements a
%filter which classifies these commands as \emph{high} or \emph{low}
%risk, and that these classifications represent the message in $\mathbf{G_{S}}$.
Let $m_{H}$ and $m_{L}$ denote messages of high and low risk, respectively,
and let $m\in M=\left\{ m_{H},m_{L}\right\} $ represent a message
in general. After $\mathcal{R}$ receives the message, it chooses an action, $a\in A=\left\{ a_{T},a_{N}\right\} $,
where $a_{T}$ represents \emph{trusting the cloud} and $a_{N}$ represents
\emph{not trusting the cloud}. 

%
%\subsubsection{Utilities, Strategies}

For the device $\mathcal{R}$, let $u_{\mathcal{R}}^{S}:\,\Theta\times M\times A\to \mathscr{U}_\mathcal{R}$, where $\mathscr{U}_\mathcal{R}\subset \mathbb{R}$.
$u_{\mathcal{R}}^{S}$ is a utility function such that $u_{\mathcal{R}}^{S}\left(\theta,m,a\right)$
gives the device's utility when the type is $\theta$, the message
is $m$, and the action is $a$. 
%The sender in the signaling game may be $\mathcal{A}$ or $\mathcal{D}$. 
Let $u_{\mathcal{A}}^{S}:\, M\times A\to\mathscr{U}_\mathcal{A}\subset \mathbb{R}$
and $u_{\mathcal{D}}^{S}:\, M\times A\to\mathscr{U}_\mathcal{D}\subset \mathbb{R}$ be utility functions
for the attacker and defender.  Note that these players only receive
utility in $\mathbf{G_{S}}$ if their own 
type controls the cloud in $\mathbf{G_{F}}$, so that type is not
longer a necessary argument for $u_{\mathcal{A}}^{S}$ and $u_{\mathcal{D}}^{S}$.

Denote the strategy of $\mathcal{R}$ by $\sigma_{\mathcal{R}}^{S}:\, A\to\left[0,1\right]$,
such that $\sigma_{\mathcal{R}}^{S}\left(a\,|\, m\right)$ gives the
mixed-strategy probability that $\mathcal{R}$ plays action $a$ when
the message is $m$. The role of the sender may be played by $\mathcal{A}$
or $\mathcal{D}$ depending on the state of the cloud, determined
by $\mathbf{G_{F}}$. Let $\sigma_{\mathcal{A}}^{S}:\, M\to\left[0,1\right]$
denote the strategy that $\mathcal{A}$ plays when she controls the
cloud, so that $\sigma_{\mathcal{A}}^{S}\left(m\right)$ gives the
probability that $\mathcal{A}$ sends message $m$. (The superscript
$S$ specifies that this strategy concerns the signaling game.) Similarly,
let $\sigma_{\mathcal{D}}^{S}:\, M\to\left[0,1\right]$ denote the
strategy played by $\mathcal{D}$ when he controls the cloud. Then
$\sigma_{\mathcal{D}}^{S}\left(m\right)$ gives the probability that
$\mathcal{D}$ sends message $m$. Let $\Gamma_{\mathcal{R}}^{S}$,
$\Gamma_{\mathcal{A}}^{S}$, and $\Gamma_{\mathcal{D}}^{S}$ denote
the sets of mixed strategies for each player.

%Next, we define mixed-strategy expected utilities. Let $\mathcal{X}\in\left\{ \mathcal{D},\mathcal{A}\right\} $ represent the cloud when it is under the control of the defender or sender. 
For $\mathcal{X}\in\left\{ \mathcal{D},\mathcal{A}\right\} $,
define functions $\bar{u}_{\mathcal{X}}^{S}:\,\Gamma_{\mathcal{R}}^{S}\times\Gamma_{\mathcal{X}}^{S}\to \mathscr{U}_\mathcal{X}$,
such that $\bar{u}_{\mathcal{X}}^{S}\left(\sigma_{\mathcal{R}}^{S},\sigma_{\mathcal{X}}^{S}\right)$
gives the expected utility to sender $\mathcal{X}$ when he or she
plays mixed-strategy $\sigma_{\mathcal{X}}^{S}$ and the receiver
plays mixed-strategy $\sigma_{\mathcal{R}}^{S}$. Equation (\ref{eq:expUtilSender})
gives $\bar{u}_{\mathcal{X}}^{S}$.

\begin{equation}
\bar{u}_{\mathcal{X}}^{S}\left(\sigma_{\mathcal{R}}^{S},\sigma_{\mathcal{X}}^{S}\right)=\underset{a\in A}{\sum}\underset{m\in M}{\sum}u_{\mathcal{X}}^{S}\left(m,a\right)\sigma_{\mathcal{R}}^{S}\left(a\,|\, m\right)\sigma_{\mathcal{X}}^{S}\left(m\right),\,\mathcal{X}\in\left\{ \mathcal{A},\mathcal{D}\right\} \label{eq:expUtilSender}
\end{equation}

Next, let $\mu:\,\Theta\to\left[0,1\right]$ represent the belief
of $\mathcal{R}$, such that $\mu\left(\theta\,|\, m\right)$ gives
the likelihood with which $\mathcal{R}$ believes that a sender who
issues message $m$ is of type $\theta$. Then define $\bar{u}_{\mathcal{R}}^{S}:\,\Gamma_{\mathcal{R}}^{S}\to\mathscr{U}_\mathcal{R}$
such that $\bar{u}_{\mathcal{R}}^{S}\left(\sigma_{\mathcal{R}}^{S}\,|\, m,\mu\left(\bullet\,|\, m\right)\right)$
gives the expected utility for $\mathcal{R}$ when it has belief $\mu$,
the message is $m$, and it plays strategy $\sigma_{\mathcal{R}}^{S}$.
$\bar{u}_{\mathcal{R}}^{S}$ is given by 
\begin{equation}
\bar{u}_{\mathcal{R}}^{S}\left(\sigma_{\mathcal{R}}^{S}\,|\, m,\mu\right)=\underset{\theta\in\Theta}{\sum}\underset{a\in A}{\sum}u_{\mathcal{R}}^{S}\left(\theta,m,a\right)\mu_{R}\left(\theta\,|\, m\right)\sigma_{\mathcal{R}}^{S}\left(a\,|\, m\right).\label{eq:expUtilReceiver}
\end{equation}

The expected utilities to the sender and receiver will determine their
incentives to control the cloud in the game $\mathbf{G_{F}}$ described in the next subsection.

\subsection{\texttt{FlipIt} Game for Cloud Control}

%In the game of \texttt{FlipIt} \cite{vanDijk2013Flip}, an attacker
%and defender vie for a resource. The attacker has the ability to completely
%and stealthily compromize the resource. The defender has the ability
%to renew his control of the resource. The attacker's and defender's
%moves both come with costs. Their utilities are determined by subtracting
%the time-averaged cost of their moves from a benefit proportional
%to the fraction of the time during which they control the resource.%

%
%\subsubsection{Time, Actions, Strategies}

The basic version of \texttt{FlipIt} \cite{vanDijk2013Flip}\footnote{See \cite{vanDijk2013Flip} for a more comprehensive definition of
the players, time, game state, and moves in \texttt{FlipIt.} Here,
we move on to describing aspects of our game important for analyzing
$\mathbf{G_{CC}}$.%
} is played in continuous time.
Assume that the defender controls the resource - here, the cloud -
at $t=0$. 
%After $t=0$, both players may move at any time. A move for the attacker consists of penetrating the cloud, and a move for the defender consists of renewing his control of the cloud. In other words, 
Moves for both players obtain control of the cloud if it is
under the other player's control.
%van Dijk and coauthors consider several classes of strategies. 
In this paper, we limit our analysis to \emph{periodic} strategies, in which
the moves of the attacker and the moves of the defender are both spaced
equally apart, and their phases are chosen randomly from a uniform
distribution. 
%Strategies can then be defined by the frequencies with
%which actions are played. 
Let $f_{\mathcal{A}}\in\mathbb{R}_{+}$ and $f_{\mathcal{D}}\in\mathbb{R}_{+}$ (where $\mathbb{R}_{+}$ represents non-negative real numbers) denote the attack and renewal
frequencies, respectively.

%\subsubsection{Utilities }
%

Players benefit from controlling the cloud, and incur costs from moving.
Let $w_{\mathcal{X}}\left(t\right)$ denote the average proportion
of the time that player $\mathcal{X}\in\left\{ \mathcal{D},\mathcal{A}\right\} $
has controlled the cloud up to time
$t$. Denote the number of moves up to $t$ per unit time of player
$\mathcal{X}$ by $z_{\mathcal{X}}\left(t\right)$. Let $\alpha_{\mathcal{D}}$
and $\alpha_{\mathcal{A}}$ represent the costs of each defender and
attacker move. In the original formulation of \texttt{FlipIt}, the authors 
consider a fixed benefit for controlling
the cloud. In our formulation, the benefit depends on the equilibrium
outcomes of the signaling game $\mathbf{G_{S}}$. 
%We formalize the equilibrium concept of our game in Section \ref{sec:SolCon}. 
Denote these equilibrium utilities of $\mathcal{D}$ and $\mathcal{A}$
by $\bar{u}_{\mathcal{D}}^{S*}$ and $\bar{u}_{\mathcal{A}}^{S*}$.
These give the expected benefit of controlling the cloud. Finally,
let $u_{\mathcal{D}}^{F}\left(t\right)$ and $u_{\mathcal{A}}^{F}\left(t\right)$
denote the time-averaged benefit of $\mathcal{D}$ and $\mathcal{A}$
up to time $t$ in $\mathbf{G_F}$. Then

\begin{equation}
u_{\mathcal{X}}^{F}\left(t\right)=\bar{u}_{\mathcal{X}}^{S*}w_{\mathcal{X}}\left(t\right)-\alpha_{\mathcal{X}}z_{\mathcal{X}}\left(t\right),\,\mathcal{X}\in\left\{ \mathcal{D},\mathcal{A}\right\},
\label{eq:FlipPayoff1}
\end{equation}
and, as time continues to evolve, the average benefits over all time
become

\begin{equation}
\underset{t\to\infty}{\liminf}\:\bar{u}_{\mathcal{X}}^{S*}w_{\mathcal{X}}\left(t\right)-\alpha_{\mathcal{X}}z_{\mathcal{X}}\left(t\right),\,\mathcal{X}\in\left\{ \mathcal{D},\mathcal{A}\right\} .
\end{equation}

We next express these expected utilities over all time as a function
of periodic strategies that $\mathcal{D}$ and $\mathcal{A}$ employ.
Let $\bar{u}_{\mathcal{X}}^{F}:\,\mathbb{R}_{+}\times\mathbb{R}_{+}\to\mathbb{R}$,
$\mathcal{X}\in\left\{ \mathcal{D},\mathcal{A}\right\} $ be expected
utility functions such that $\bar{u}_{\mathcal{D}}^{F}\left(f_{\mathcal{D}},f_{\mathcal{A}}\right)$
and $\bar{u}_{\mathcal{A}}^{F}\left(f_{\mathcal{D}},f_{\mathcal{A}}\right)$
give the average utility to $\mathcal{D}$ and $\mathcal{A}$, respectively,
when they play with frequencies $f_{\mathcal{D}}$ and $f_{\mathcal{A}}$.
If $f_{\mathcal{D}}\geq f_{\mathcal{A}} > 0$, it can be shown that 
\begin{equation}
\bar{u}_{\mathcal{D}}^{F}\left(f_{\mathcal{D}},f_{\mathcal{A}}\right)=\bar{u}_{\mathcal{D}}^{S*}\left(1-\frac{f_{\mathcal{A}}}{2f_{\mathcal{D}}}\right)-\alpha_{\mathcal{D}}f_{\mathcal{D}},\label{eq:expUtilDefFlip-DFaster}
\end{equation}

\begin{equation}
\bar{u}_{\mathcal{A}}^{F}\left(f_{\mathcal{D}},f_{\mathcal{A}}\right)=\bar{u}_{\mathcal{A}}^{S*}\frac{f_{\mathcal{A}}}{2f_{\mathcal{D}}}-\alpha_{\mathcal{A}}f_{\mathcal{A}},\label{eq:expUtilAtkFlip-DFaster}
\end{equation}
while if $0 \leq f_{\mathcal{D}} < f_{\mathcal{A}}$, then
\begin{equation}
\bar{u}_{\mathcal{D}}^{F}\left(f_{\mathcal{D}},f_{\mathcal{A}}\right)=\bar{u}_{\mathcal{D}}^{S*}\frac{f_{\mathcal{D}}}{2f_{\mathcal{A}}}-\alpha_{\mathcal{D}}f_{\mathcal{D}},\label{eq:expUtilDefFlip-AFaster}
\end{equation}

\begin{equation}
\bar{u}_{\mathcal{A}}^{F}\left(f_{\mathcal{D}},f_{\mathcal{A}}\right)=\bar{u}_{\mathcal{A}}^{S*}\left(1-\frac{f_{\mathcal{D}}}{2f_{\mathcal{A}}}\right)-\alpha_{\mathcal{A}}f_{\mathcal{A}},\label{eq:expUtilAtkFlip-AFaster}
\end{equation}

and if $f_{\mathcal{A} }=0$, we have 
\begin{equation}
\bar{u}_{\mathcal{A}}^{F}\left(f_{\mathcal{D}},f_{\mathcal{A}}\right)= 0, \, \, \, \,  \bar{u}_{\mathcal{D}}^{F}\left(f_{\mathcal{D}},f_{\mathcal{A}}\right)= \bar{u}_\mathcal{D}^{S*}-\alpha_\mathcal{D}f_\mathcal{D}.
\label{eq:expUtilAttNo}
\end{equation}
%
%\begin{equation}
%\bar{u}_{\mathcal{D}}^{F}\left(f_{\mathcal{D}},f_{\mathcal{A}}\right)= 1
%\label{eq:expUtilDeffAttNo}
%\end{equation}

Equations (\ref{eq:expUtilDefFlip-DFaster})-(\ref{eq:expUtilAttNo}) with Equation (\ref{eq:expUtilSender}) for $\bar{u}_{\mathcal{X}}^{S}$,
$\mathcal{X\in\left\{ \mathcal{D},\mathcal{A}\right\} }$ and Equation
(\ref{eq:expUtilReceiver}) for $\bar{u}_{\mathcal{R}}^{S}$ will
be main ingredients in our equilibrium concept in the next section. 
%%%%%%%%%%%%%%%%%%%%%%%%%%%%%%%%%%%%%%
\section{Solution Concept}
\label{sec:SolCon}
In this section, we develop a new equilibrium concept for our \texttt{CloudControl}
game $\mathbf{G{}_{CC}}$. We study the equilibria of the \texttt{FlipIt}
and signaling games individually, and then show how they can be related
through a fixed-point equation in order to obtain an overall equilibrium
for $\mathbf{G_{CC}}.$

\subsection{Signaling Game Equilibrium}
\label{sub:SigAna}

Signaling games are a class of dynamic Bayesian games. 
%We can describe equilibrium behavior in signaling games using the concept of \emph{Bayesian equilibrium}, which requires that each player maximizes his or her utility given the type of the player. In addition, the equilibrium refinement of \emph{perfect Bayesian equilibrium} (PBE) in signaling games requires that beliefs are updated in a Bayesian manner when it is possible. 
Applying the concept of \emph{perfect Bayesian equilibrium}  (as it \emph{e.g.},  \cite{fudenberg1991game}) to $\mathbf{G_{S}}$, we have Definition \ref{def:PBE}. 

\begin{defn} 
\begin{flushleft}
\label{def:PBE}Let the functions $\bar{u}_{\mathcal{X}}^{S}\left(\sigma_{\mathcal{R}}^{S},\sigma_{\mathcal{X}}^{S}\right),\,\mathcal{X}\in\left\{ \mathcal{D},\mathcal{A}\right\} $
and $\bar{u}_{\mathcal{R}}^{S}\left(\sigma_{\mathcal{R}}^{S}\right)$
be formulated according to Equation (\ref{eq:expUtilSender}) and
Equation (\ref{eq:expUtilReceiver}), respectively. Then a \emph{perfect Bayesian equilibrium} of the signaling
game $\mathbf{G_{S}}$ is a strategy profile $\left(\sigma_{\mathcal{D}}^{S*},\sigma_{\mathcal{A}}^{S*},\sigma_{\mathcal{R}}^{S*}\right)$
and posterior beliefs $\mu\left(\bullet\,|\, m\right)$ such that
\begin{equation}
\forall\mathcal{X}\in\left\{ \mathcal{D},\mathcal{A}\right\} ,\,\sigma_{\mathcal{X}}^{S*}\left(\bullet\right)\in\underset{\sigma_{\mathcal{X}}^{S}}{\arg\max}\:\bar{u}_{\mathcal{X}}^{S}\left(\sigma_{\mathcal{R}}^{S*},\sigma_{\mathcal{X}}^{S}\right),
\end{equation}
\begin{equation}
\forall m\in M,\,\sigma_{\mathcal{R}}^{S*}\left(\bullet\,|\, m\right)\in\underset{\sigma_{\mathcal{R}}^{S}}{\arg\max}\:\bar{u}_{\mathcal{R}}^{S}\left(\sigma_{\mathcal{R}}^{S}\,|\, m,\mu\left(\bullet\,|\, m\right)\right),
\end{equation}
\begin{equation}
\mu\left(\theta\,|\, m\right)=\frac{1\left\{ \theta=\theta_{\mathcal{A}}\right\} \sigma_{\mathcal{A}}^{S*}\left(m\right)p+1\left\{ \theta=\theta_{\mathcal{D}}\right\} \sigma_{\mathcal{D}}^{S*}\left(m\right)\left(1-p\right)}{\sigma_{\mathcal{A}}^{S*}\left(m\right)p+\sigma_{\mathcal{D}}^{S*}\left(m\right)\left(1-p\right)},
\end{equation}
if $\sigma_{\mathcal{A}}^{S*}\left(m\right)p+\sigma_{\mathcal{D}}^{S*}\left(m\right)\left(1-p\right)\neq0$,
and 
\begin{equation}
\mu\left(\theta\,|\, m\right)=\text{any distribution on }\Theta,
\end{equation}
if $\sigma_{\mathcal{A}}^{S*}\left(m\right)p+\sigma_{\mathcal{D}}^{S*}\left(m\right)\left(1-p\right)=0$. 

\par\end{flushleft}

\end{defn} 

Next, let $\bar{u}_{\mathcal{D}}^{S*}$, $\bar{u}_{\mathcal{A}}^{S*}$,
and $\bar{u}_{\mathcal{R}}^{S*}$ be the utilities for the defender,
attacker, and device, respectively, when they play according to a
strategy profile $\left(\sigma_{\mathcal{D}}^{S*},\sigma_{\mathcal{A}}^{S*},\sigma_{\mathcal{R}}^{S*}\right)$
and belief $\mu\left(\bullet\,|\, m\right)$ that satisfy the conditions
for a perfect Bayesian equilibrium. 
%The equilibrium strategies and utilities depend on the prior probability of the types $\theta_{\mathcal{A}}$ and $\theta_{\mathcal{D}}$, which are given by $p$ and $1-p$, respectively. In some cases, a single $p$ can map to more than one pair set of equilibrium utilities. 
Define a set-valued mapping $T^{S}:\,\left[0,1\right]\to2^{\mathcal{U_{D}}\times\mathcal{U}_{A}}$ such that
$T^S\left(p;G_S\right)$ gives the set of equilibrium utilities of the defender and attacker when the prior probabilities are $p$ and $1-p$ and the signaling game utilities are parameterized by $G_S$\footnote{Since $\mathcal{R}$ does not take part in $\mathbf{G_{S}}$, it is
not necessary to include $\bar{u}_{\mathcal{R}}^{S*}$ as an output
of the mapping. %
}. We have 
\begin{equation}
\left\{ \left(\bar{u}_{\mathcal{D}}^{S*},\bar{u}_{\mathcal{A}}^{S*}\right)\right\} =T^S\left(p;G_S\right).\label{eq:Ts}
\end{equation}
We will employ $T^{S}$ as part of the definition of an overall equilibrium
for $\mathbf{G_{CC}}$ after examining the equilibrium of the \texttt{FlipIt}
game.

\subsection{\texttt{FlipIt} Game Equilibrium}
 
%when the attacker and defender's strategies are limited to periodic strategies, can be considered as a static game. \emph{Nash equilibrium} (NE) is an adequate criteria for this class of games. In an NE, each player selects an optimal strategy given the other player's strategy choice. 
The appropriate equilibrium concept for the \texttt{FlipIt} game, when $\mathcal{A}$ and $\mathcal{D}$ are restricted to periodic strategies, is \emph{Nash equilibrium} \cite{nash1950equilibrium}. Definition \ref{def:Nash} applies the concept of Nash Equilibrim to $\mathbf{G_F}$. 
\begin{defn} \label{def:Nash}
A \emph{Nash equilibrium} of the game $\mathbf{G_{F}}$ is a strategy
profile $\left(f_{\mathcal{D}}^{*},f_{\mathcal{A}}^{*}\right)$ such
that 
\begin{equation}
f_{\mathcal{D}}^{*}\in\underset{f_{\mathcal{D}}}{\arg\max}\:\bar{u}_{\mathcal{D}}^{F}\left(f_{\mathcal{D}},f_{\mathcal{A}}^{*}\right),
\end{equation}
\begin{equation}
f_{\mathcal{A}}^{*}\in\underset{f_{\mathcal{A}}}{\arg\max}\:\bar{u}_{\mathcal{D}}^{F}\left(f_{\mathcal{D}}^{*},f_{\mathcal{A}}\right),
\end{equation}
where $\bar{u}_{\mathcal{D}}^{F}$ and $\bar{u}_{\mathcal{A}}^{F}$
are computed by Equation (\ref{eq:expUtilDefFlip-DFaster}) and Equation
(\ref{eq:expUtilAtkFlip-DFaster}) if $f_{\mathcal{D}}\geq f_{\mathcal{A}}$
and Equation (\ref{eq:expUtilDefFlip-AFaster}) and Equation (\ref{eq:expUtilAtkFlip-AFaster})
if $f_{\mathcal{D}}\leq f_{\mathcal{A}}$. \end{defn} 

To find an overall equilibrium of $\mathbf{G_{CC}}$, we are interested
in the proportion of time that $\mathcal{A}$ and $\mathcal{D}$ control
the cloud. As before, denote these proportions by $p$ and $1-p$,
respectively. These proportions (as in \cite{bowers2012defending}) can be found from the equilibrium frequencies by 

\begin{equation}
p=\begin{cases}
0,\text{ if } & f_{\mathcal{A}}=0\\
\frac{f_{\mathcal{A}}}{2f_{\mathcal{D}}},\text{ if } & f_{\mathcal{D}}\geq f_{\mathcal{A}}>0\\
1-\frac{f_{\mathcal{D}}}{2f_{\mathcal{A}}},\text{ if } & f_{\mathcal{A}}>f_{\mathcal{D}}\geq0
\end{cases}.\label{eq:p}
\end{equation}

Let $G_F$ parameterize the \texttt{FlipIt} game.
Now, we can define a mapping $T^{F}:\,\mathcal{U_{D}}\times\mathcal{U_{A}}\to\left[0,1\right]$
such that the expression $T^{F}\left(\bar{u}_{\mathcal{D}}^{S*},\bar{u}_{\mathcal{A}}^{S*};G_F\right)$ gives the proportion of time that the attacker controls the cloud in equilibrium from the values of controlling the cloud for the defender and the attacker. This mapping gives

\begin{equation}
p=T^{F}\left(\bar{u}_{\mathcal{D}}^{S*},\bar{u}_{\mathcal{A}}^{S*};G_F\right).\label{eq:Tf}
\end{equation}

In addition to interpreting $p$ as the proportion of time that the
attacker controls the cloud, we can view it as the likelihood that,
at any random time, the cloud will be controlled by the attacker.
Of course, this is precisely the value $p$ of interest in $\mathbf{G_{S}}$.
Clearly, $\mathbf{G_{F}}$ and $\mathbf{G_{S}}$ are coupled by Equations
(\ref{eq:Ts}) and (\ref{eq:Tf}). These two equations specify the overall
equilibrium for the \texttt{CloudControl} game $\mathbf{G_{CC}}$
through a fixed-point equation, which we describe next.

\subsection{Gestalt Equilibrium of $\mathbf{G_{CC}}$}

\begin{figure}[t]
\begin{center}
\includegraphics[width=0.5\columnwidth]{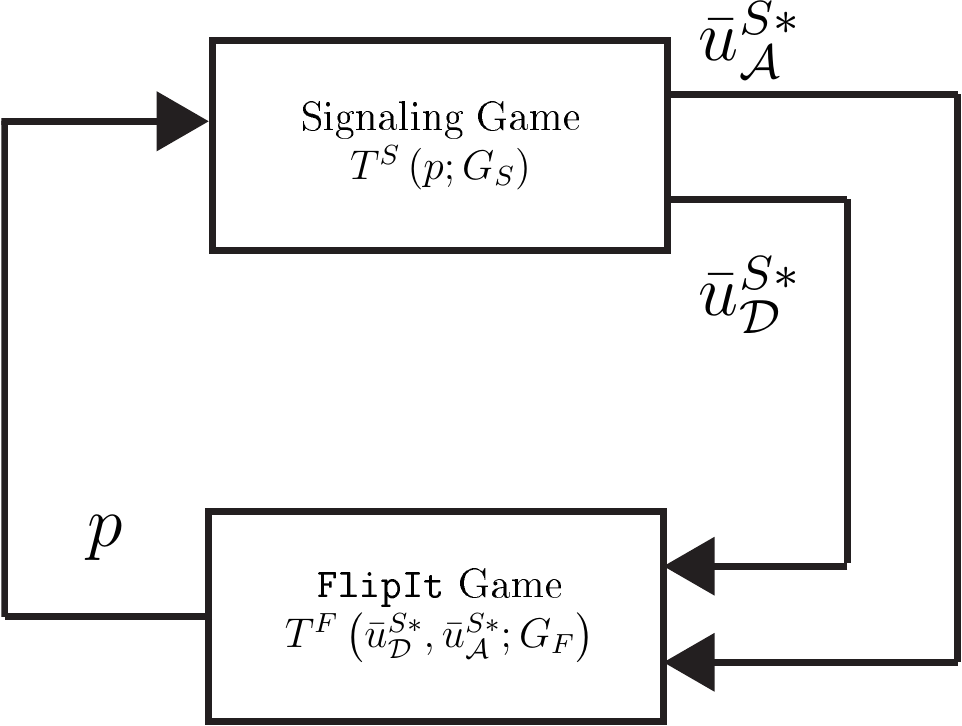}

\caption{$\mathbf{G_S}$ and $\mathbf{G_F}$ interact because the utilities in the \texttt{FlipIt} game are derived from the output of the signaling game, and the output of the \texttt{FlipIt} game is used to define prior probabilities in the signaling game. We call the fixed-point of the composition of these two relationships a Gestalt equilibrium.}
\label{fig:Solconcept}
\end{center}
\end{figure}

When the \texttt{CloudControl} game $\mathbf{G_{CC}}$ is in equilibrium
the mapping from the parameters of $\mathbf{G_{S}}$ to that game's
equilibrium and the mapping from the parameters of $\mathbf{G_{F}}$
to that game's equilibrium are simultaneously satisfied as shown in Fig. \ref{fig:Solconcept}. Definition \ref{thm:fixedPoint}
formalizes this equilibrium, which we call \textit{Gestalt equilibrium}. 

\newtheorem{Def}{Definition}

\begin{defn} \label{thm:fixedPoint} (\textbf{Gestalt equilibrium}) The cloud control ratio
 $p^{\dagger}\in\left[0,1\right]$ and equilibrium signaling game utilities
 $\bar{u}_{\mathcal{D}}^{S\dagger}$ and $\bar{u}_{\mathcal{A}}^{S\dagger}$
constitute a Gestalt equilibrium of the game $\mathbf{G_{CC}}$
composed of coupled games $\mathbf{G_{S}}$ and $\mathbf{G_{F}}$
if the two components of Equation (\ref{eq:TsEquilib}) are simultaneously
satisfied. 
\begin{equation}
\left(\bar{u}_{\mathcal{D}}^{S\dagger},\bar{u}_{\mathcal{A}}^{S\dagger}\right)\in T^{S}\left(p^{\dagger};G_S\right), \, \, \, \, \, p^{\dagger}=T^{F}\left(\bar{u}_{\mathcal{D}}^{S\dagger},\bar{u}_{\mathcal{A}}^{S\dagger};G_F\right)
\label{eq:TsEquilib}
\end{equation}

%
%\begin{equation}
%p^{\dagger}=T^{F}\left(\bar{u}_{\mathcal{D}}^{S\dagger},\bar{u}_{\mathcal{A}}^{S\dagger}\right)\label{eq:TfEquilib}
%\end{equation}

In short, the signaling game utilities $\left(\bar{u}_{\mathcal{D}}^{S\dagger},\bar{u}_{\mathcal{A}}^{S\dagger}\right)$ must satisfy the fixed-point
equation
\begin{equation}
\left(\bar{u}_{\mathcal{D}}^{S\dagger},\bar{u}_{\mathcal{A}}^{S\dagger}\right)\in
T^S\left(T^F\left(\bar{u}_{\mathcal{D}}^{S\dagger},\bar{u}_{\mathcal{A}}^{S\dagger};G_F\right);G_S\right).
\end{equation}

In this equilibrium, $\mathcal{A}$ receives $\bar{u}_{\mathcal{A}}^{F}$ 
according to Equation (\ref{eq:expUtilAtkFlip-DFaster}), Equation 
(\ref{eq:expUtilAtkFlip-AFaster}), or Equation (\ref{eq:expUtilAttNo}),  $\mathcal{D}$ receives 
$\bar{u}_{\mathcal{D}}^{F}$ 
according to Equation (\ref{eq:expUtilDefFlip-DFaster}), Equation
 (\ref{eq:expUtilDefFlip-AFaster}), or Equation (\ref{eq:expUtilAttNo}),  and $\mathcal{R}$ receives 
$\bar{u}_{\mathcal{R}}^{S}$ according to Equation (\ref{eq:expUtilReceiver}).

\end{defn} 

Solving for the equilibrium of $\mathbf{G_{CC}}$ requires
a fixed-point equation essentially because the games $\mathbf{G_{F}}$
and $\mathbf{G_{S}}$ are played according to \emph{prior committment}.
Prior commitment specifies that players in $\mathbf{G_{S}}$ do not
know the outcome of $\mathbf{G_{F}}$. This structure prohibits us
from using a sequential concept such as sub-game perfection and suggests
instead a fixed-point equation.

%Having described the structure of the \texttt{CloudControl} game and specified the conditions necessary for a flip-fixed equilibrium, we now solve for the equilibrium in various cases.

%%%%%%%%%%%%%%%%%%%%%%%%%%%%%%%%%%%%%% 
\section{Analysis}
\label{sec:Ana}
In this section, we analyze the game proposed in Section \ref{sec:SysMod}
based on our solution concept in Section \ref{sec:SolCon}. First,
we analyze the signaling game and calculate the corresponding equilibria.
Then, we solve the $\mathtt{FlipIt}$ game for different values of
expected payoffs resulting from signaling game. Finally, we describe the solution of the combined game.

\subsection{Signaling Game Analysis}\label{ssec:sigAna}

The premise of $\mathbf{G_{CC}}$ allows us to make some basic assumptions
about the utility parameters that simplifies the search for equilibria.
We expect these assumptions to be true across many different contexts.
\begin{description}
\item[A1)]$u_{\mathcal{R}}(\theta_{\mathcal{D}},m_{L},a_{T})>u_{\mathcal{R}}(\theta_{\mathcal{D}},m_{L},a_{N})$:
It is beneficial for the receiver to trust a low risk message from the defender.
\item[A2)]$u_{\mathcal{R}}(\theta_{\mathcal{A}},m_{H},a_{T})<u_{\mathcal{R}}(\theta_{\mathcal{A}},m_{H},a_{N})$:
It is harmful for the receiver to trust a high risk message from the attacker.
\item[A3)]$\forall{m,m'}\in{M},\;
u_{\mathcal{A}}(m,a_{T})>u_{\mathcal{A}}(m',a_{N})$ and 
$\forall{m,m'}\in{M}\;,
u_{\mathcal{D}}(m,a_{T})>u_{\mathcal{D}}(m',a_{N})$: 
Both types of sender prefer that either of their messages is trusted rather than that either of their messages is rejected.
\item[A4)]$u_{\mathcal{A}}(m_{H},a_{T})>u_{\mathcal{A}}(m_{L},a_{T})$:
The attacker prefers an outcome in which the receiver trusts his high risk message to an outcome in which the receiver trusts his low risk message.
\end{description}
%Based on these assumptions, we can search for separating and pooling%equilibria. (The derivations for these equilibria are given in Appendix \ref{SepDLAH} and Appendix \ref{PoolL}.) First, consider pooling equilibria. In a pooling equilibrium,%the receiver relies only on prior probability (here $p$ and $1-p$)%to form its beliefs. Based on these prior probability, it maximizes%its expected utility by choosing $a_{T}$ or $a_{N}$ given the message%that it observes. Define \emph{trust benefits} for high and low messages%when the prior probability of $\theta=\theta_{\mathcal{A}}$ is $p$%by $TB_{H}\left(p\right)$ and $TB_{L}\left(p\right)$ in Equations%(\ref{eq:tbH}) and (\ref{eq:tbL}).

Pooling equilibria of the signaling game differ depending on the prior
probabilities $p$ and $p-1$.  Specifically, the messages on which $\mathcal{A}$ 
and $\mathcal{D}$ pool and the equilibrium action of $\mathcal{R}$ depend
on quantities in Equations
(\ref{eq:tbH}) and (\ref{eq:tbL}) which we call \textit{trust benefits.}

\begin{equation}
TB_{H}\left(p\right)=\begin{array}{c}
p\left[u_{\mathcal{R}}\left(\theta_{\mathcal{A}},m_{H},a_{T}\right)-u_{\mathcal{R}}\left(\theta_{\mathcal{A}},m_{H},a_{N}\right)\right]\\
+\left(1-p\right)\left[u_{\mathcal{R}}\left(\theta_{\mathcal{D}},m_{H},a_{T}\right)-u_{\mathcal{R}}\left(\theta_{\mathcal{D}},m_{H},a_{N}\right)\right]
\end{array}\label{eq:tbH}
\end{equation}
\begin{equation}
TB_{L}\left(p\right)=\begin{array}{c}
p\left[u_{\mathcal{R}}\left(\theta_{\mathcal{A}},m_{L},a_{T}\right)-u_{\mathcal{R}}\left(\theta_{\mathcal{A}},m_{L},a_{N}\right)\right]\\
+\left(1-p\right)\left[u_{\mathcal{R}}\left(\theta_{\mathcal{D}},m_{L},a_{T}\right)-u_{\mathcal{R}}\left(\theta_{\mathcal{D}},m_{L},a_{N}\right)\right]
\end{array}\label{eq:tbL}
\end{equation}

$TB_{H}\left(p\right)$ and $TB_{L}\left(p\right)$ give the benefit
of trusting (compared to not trusting) high and low messages, respectively,
when the prior probability is $p$. These quantities specify whether
$\mathcal{R}$ will trust a message that it receives in a pooling
equilibrium. If $TB_H\left(p\right)$ (respectively, $TB_L\left(p\right)$) is positive, then, in equilibrium, $\mathcal{R}$ will trust all messages when the senders pool on $m_H$ (respectively, $m_L$).

\begin{figure}[t]
\begin{centering}
\includegraphics[width=0.6\columnwidth]{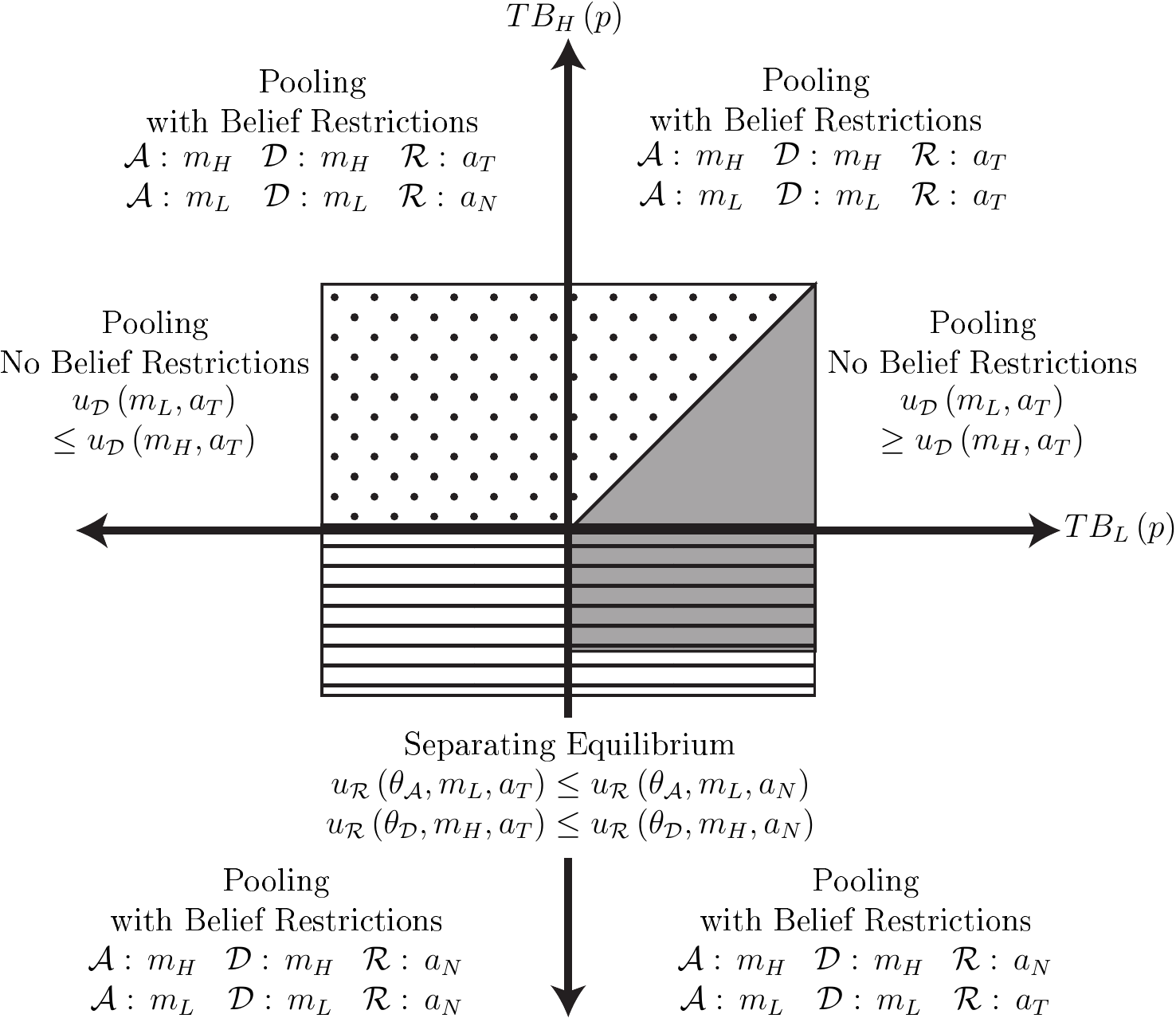}
\par\end{centering}

\begin{centering}
\protect\caption{The four quadrants represent parameter regions of $\mathbf{G_S}$.  The regions vary based on the tpyes of pooling equilibria that they support.  For instance, quadrant IV supports a pooling equilibrium in which $\mathcal{A}$ and $\mathcal{D}$ both send $m_H$ and $\mathcal{R}$ plays $a_N$, as well as a pooling equilibrium in which $\mathcal{A}$ and $\mathcal{D}$ both send $m_L$ and $\mathcal{R}$ plays $a_T$.  The shaded regions denote special equilibria that occur under further parameter restrictions.}

\par\end{centering}

\centering{}\label{fig:equilibRegions} 
\end{figure}

%When $TB_{H}\left(p\right)>0$ ($<0$), $\mathcal{R}$ will trust (not trust, respectively) $m_{H}$, and when $TB_{L}\left(p\right)>0$ ($<0$), $\mathcal{R}$ will trust (not trust, respectively) $m_{L}$. 
We illustrate the different possible combinations of $TB_{H}\left(p\right)$
and $TB_{L}\left(p\right)$ in the quadrants of Fig. \ref{fig:equilibRegions}.
The labeled messages and actions for the sender and receiver, respectively,
in each quadrant denote these pooling equilibria. These pooling equilibria apply throughout each entire
quadrant. Note that we have not listed the requirements on belief
$\mu$ here. These are addressed in the Appendix \ref{ssec:poolEq}, and become especially
important for various equilibrium refinement procedures.

The shaded regions of Fig. \ref{fig:equilibRegions} denote additional
special equilibria which only occur under the additional parameter
constraints listed within the regions. (The geometrical shapes of the shaded regions are not meaningful, but their overlap and location relative to the four quadrants are accurate.)
%These equilibria are listed in table \ref{tab:specialEq}. 
The dotted and uniformly shaded zones contain equilibria similar to those already denoted in
the equilibria for each quadrant, except that they do not require
restrictions on $\mu$. The zone with horizontal bars denotes 
the game's only separating equilibrium. It is a rather unproductive one for $\mathcal{D}$ and $\mathcal{A}$,
since their messages are not trusted. (See the derivation in Appendix \ref{ssec:sepEq}.)
%\begin{table*}%\protect\caption{Special Equilibria\label{tab:specialEq}}%\centering{}%%\begin{tabular}{|c|c|c|c|}%\hline %Equilibrium Name & Required Constraints & Equilibrium Actions & Note\tabularnewline%\hline %\hline %S1 (red) & fill in & $\mathcal{A}:\, m_{H}$ $\mathcal{D}:\, m_{H}$ $\mathcal{R}:\, a_{T}$ & $\mu\in\left[0,1\right]$ \tabularnewline%\hline %S2 (blue) & fill in & $\mathcal{A}:\, m_{L}$ $\mathcal{D}:\, m_{L}$ $\mathcal{R}:\, a_{T}$ & $\mu\in\left[0,1\right]$ \tabularnewline%\hline %S3 (yellow) & fill in & $\mathcal{A}:\, m_{L}$ $\mathcal{D}:\, m_{H}$ $\mathcal{R}\text{ for }m_{L},m_{H}:\, a_{N},a_{N}$ & separating\tabularnewline%\hline %\end{tabular}%\end{table*}
The equilibria depicted in Fig. \ref{fig:equilibRegions}
will become the basis of analyzing the mapping $T^{S}\left(p;G_S\right)$,
which will be crucial for forming our fixed-point equation that defines
the Gestalt equilibrium. Before studying this mapping, however, we
first analyze the equilibria of the \texttt{FlipIt} game on its own.

\subsection{\texttt{FlipIt} Analysis}

\label{sub:FlipAna}

In this subsection, we calculate the Nash equilibrium in the \texttt{FlipIt} game. Equations (\ref{eq:expUtilDefFlip-DFaster})-(\ref{eq:expUtilAttNo})
represent both players' utilities in \texttt{FlipIt} game. The
solution of this game is similar to what has presented in \cite{vanDijk2013Flip,bowers2012defending}, except that the reward of controlling
the resource may vary. To calculate Nash equilibrium, we
normalize both players' benefit with respect to the reward of controlling
the resource. %Player $\mathcal{D}$'s best response, i.e., $BR_{\mathcal{D}}(f_{\mathcal{A}})$, is as follows:%\begin{equation}%BR_{\mathcal{D}}(f_{\mathcal{A}})=\begin{cases}%\sqrt{\dfrac{f_{\mathcal{A}}\bar{u}_{\mathcal{D}}^{S*}}{2\alpha_{\mathcal{D}}}} & \quad f_{\mathcal{A}}<\dfrac{\bar{u}_{\mathcal{D}}^{S*}}{2\alpha_{\mathcal{D}}}\\%\bigg[0,\sqrt{\dfrac{f_{\mathcal{A}}\bar{u}_{\mathcal{D}}^{S*}}{2\alpha_{\mathcal{D}}}}\bigg] & \quad f_{\mathcal{A}}=\dfrac{\bar{u}_{\mathcal{D}}^{S*}}{2\alpha_{\mathcal{D}}}\\%0 & \quad f_{\mathcal{A}}>\dfrac{\bar{u}_{\mathcal{D}}^{S*}}{2\alpha_{\mathcal{D}}}%\end{cases}\label{eq:DefBestR}%\end{equation}%Player $\mathcal{A}$'s best response, i.e., $BR_{\mathcal{A}}(f_{\mathcal{D}})$,%is as follows:%\begin{equation}%BR_{\mathcal{A}}(f_{\mathcal{D}})=\begin{cases}%\sqrt{\dfrac{f_{\mathcal{D}}\bar{u}_{\mathcal{A}}^{S*}}{2\alpha_{\mathcal{A}}}} & \quad f_{\mathcal{D}}<\dfrac{\bar{u}_{\mathcal{A}}^{S*}}{2\alpha_{\mathcal{A}}}\\%\bigg[0,\sqrt{\dfrac{f_{\mathcal{D}}\bar{u}_{\mathcal{A}}^{S*}}{2\alpha_{\mathcal{A}}}}\bigg] & \quad f_{\mathcal{D}}=\dfrac{\bar{u}_{\mathcal{A}}^{S*}}{2\alpha_{\mathcal{A}}}\\%0 & \quad f_{\mathcal{D}}>\dfrac{\bar{u}_{\mathcal{A}}^{S*}}{2\alpha_{\mathcal{A}}}%\end{cases}\label{eq:AttBestR}%\end{equation}%The Nash equilibrium is the intersection of both players' best responses.
For different cases, the frequencies of move at Nash equilibrium are:

$\bullet\,\,\,\dfrac{\alpha_{\mathcal{D}}}{{\bar{u}_{\mathcal{D}}^{S*}}}<\dfrac{\alpha_{\mathcal{A}}}{{\bar{u}_{\mathcal{A}}^{S*}}}$
and ${\bar{u}_{\mathcal{A}}^{S*}},{\bar{u}_{\mathcal{D}}^{S*}}>0$:

\begin{equation}
f_{\mathcal{D}}^{*}=\dfrac{{\bar{u}_{\mathcal{A}}^{S*}}}{2\alpha_{\mathcal{A}}},\,\, f_{\mathcal{A}}^{*}=\dfrac{\alpha_{\mathcal{D}}}{2\alpha_{\mathcal{A}}^{2}}\times\dfrac{({\bar{u}_{\mathcal{A}}^{S*}})^{2}}{{\bar{u}_{\mathcal{D}}^{S*}}}\label{eq:FlipNE1},
\end{equation}

$\bullet\,\,\,\dfrac{\alpha_{\mathcal{D}}}{{\bar{u}_{\mathcal{D}}^{S*}}}>\dfrac{\alpha_{\mathcal{A}}}{{\bar{u}_{\mathcal{A}}^{S*}}}$
and ${\bar{u}_{\mathcal{A}}^{S*}},{\bar{u}_{\mathcal{D}}^{S*}}>0$:

\begin{equation}
f_{\mathcal{D}}^{*}=\dfrac{\alpha_{\mathcal{A}}}{2\alpha_{\mathcal{D}}^{2}}\times\dfrac{({\bar{u}_{\mathcal{D}}^{S*}})^{2}}{{\bar{u}_{\mathcal{A}}^{S*}}},\,\, f_{\mathcal{A}}^{*}=\dfrac{{\bar{u}_{\mathcal{D}}^{S*}}}{2\alpha_{\mathcal{D}}}\label{eq:FlipNE2},
\end{equation}

$\bullet\,\,\,\dfrac{\alpha_{\mathcal{D}}}{{\bar{u}_{\mathcal{D}}^{S*}}}=\dfrac{\alpha_{\mathcal{A}}}{{\bar{u}_{\mathcal{A}}^{S*}}}$
and ${\bar{u}_{\mathcal{A}}^{S*}},{\bar{u}_{\mathcal{D}}^{S*}}>0$:

\begin{equation}
f_{\mathcal{D}}^{*}=\dfrac{{\bar{u}_{\mathcal{A}}^{S*}}}{2\alpha_{\mathcal{A}}},\,\, f_{\mathcal{A}}^{*}=\dfrac{{\bar{u}_{\mathcal{D}}^{S*}}}{2\alpha_{\mathcal{D}}}\label{eq:FlipNE3},
\end{equation}

$\bullet\,\,\,{\bar{u}_{\mathcal{A}}^{S*}}\leq0$:

\begin{equation}
f_{\mathcal{D}}^{*}=f_{\mathcal{A}}^{*}=0\label{eq:FlipNE4},
\end{equation}

$\bullet\,\,\,{\bar{u}_{\mathcal{A}}^{S*}}>0$ and ${\bar{u}_{\mathcal{D}}^{S*}}\leq0$:

\begin{equation}
f_{\mathcal{D}}^{*}=0\,\,\,\,\, f_{\mathcal{A}}^{*}=0^{+}\label{eq:FlipNE5}.
\end{equation}

In the case that $\bar{u}_{\mathcal{A}}^{S*}\leq0$, the attacker has
no incentive to attack the cloud.  In this case, the defender need not move since
we assume that she controls the cloud initially.  In the case that 
$\bar{u}_{\mathcal{A}}^{S*}>0$ and ${\bar{u}_{\mathcal{D}}^{S*}}\leq0$,
only the attacker has an incentive to control the cloud.  We use
 $f_{\mathcal{A}}^{*}=0^{+}$ to signify that the attacker moves 
only once.  Since the defender
never moves, the attacker's single move is enough to retain control of the
cloud at all times.

Next, we put together the analysis of $\mathbf{G_{S}}$ and $\mathbf{G_{F}}$
in order to study the Gestalt equilibria of the entire game.

\subsection{$\mathbf{G_{CC}}$ Analysis}

To identify the Gestalt Equilibrium of $\mathbf{G_{CC}}$, it is necessary to examine the mapping $T^S\left(p;G_S\right)$ for all $p\in{\left[0,1\right]}$. As noted in Section \ref{ssec:sigAna}, this mapping depends on $TB_H\left(p\right)$ and $TB_L\left(p\right)$.
From assumptions A1-A4, it is possible to verify that $\left(TB_{L}\left(0\right),TB_{H}\left(0\right)\right)$
must fall in Quadrant I or Quadrant IV and that $\left(TB_{L}\left(1\right),TB_{H}\left(1\right)\right)$
must lie in Quadrant III or Quadrant IV. There are numerous
ways in which the set $\left(TB_{L}\left(p\right),TB_{H}\left(p\right)\right),\, p\in\left[0,1\right]$
can transverse different parameter regions. Rather than enumerating
all of them, we consider one here.

\begin{figure}
\begin{centering}
\includegraphics[width=0.6\columnwidth]{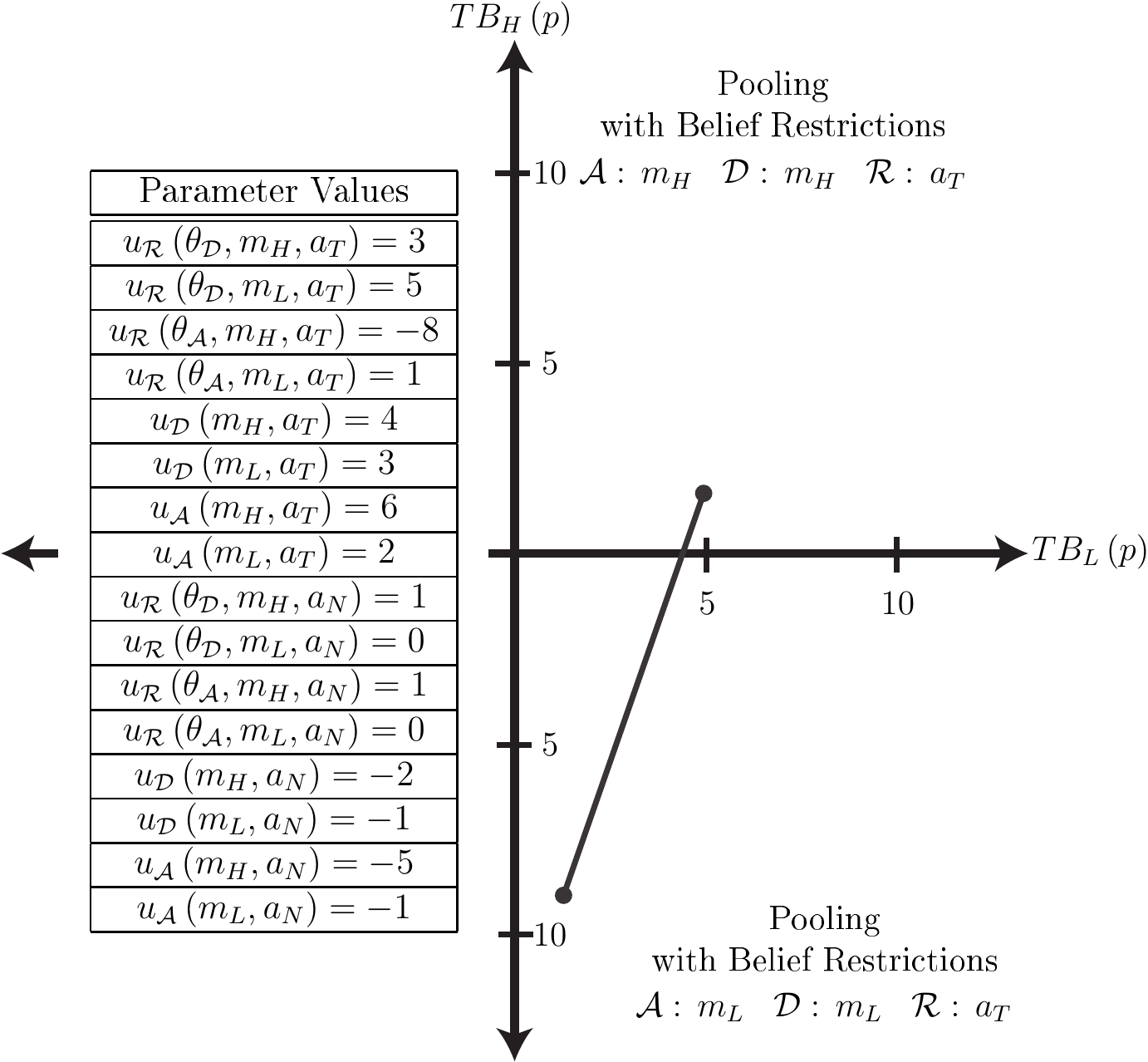}\label{fig:pathParams}\protect\caption{For the parameter values overlayed on the figure, as $p$ ranges from $0$ to $1$, $TB_H\left(p\right)$ and $TB_L\left(p\right)$ move from Quadrant I to Quadrant IV.  The equilibria supported in each of these quadrants, as well as the equilibria supported on the interface between them, are presented in Table \ref{tab:TsEqQIQIV}.}

\par\end{centering}

\end{figure}

Consider parameters such that $TB_{L}\left(0\right),TB_{H}\left(0\right)>0$
and $TB_{L}\left(1\right)>0$ but $TB_{H}\left(1\right)<0$\footnote{These parameters must satisfy $u_{\mathcal{R}}\left(\theta_{\mathcal{D}},m_{H},a_{T}\right)>u_{\mathcal{R}}\left(\theta_{\mathcal{D}},m_{H},a_{N}\right)$ 
and  $u_{\mathcal{R}}\left(\theta_{\mathcal{A}},m_{L},a_{T}\right)>u_{\mathcal{R}}\left(\theta_{\mathcal{A}},m_{L},a_{N}\right)$.
Here, we give them specific values in order to plot the data.%
}. This leads to an $\mathscr{L}$ that will traverse from Quadrant
I to Quadrant IV. Let us also assume that $u_{\mathcal{D}}\left(m_{L},a_{T}\right)<u_{\mathcal{D}}\left(m_{H},a_{T}\right)$,
so that Equilibrium 5 is not feasible. 
In Fig. \ref{fig:pathParams}, we give specific values of parameters that satisfy these conditions, and we plot $\left(TB_L\left(p\right),TB_H\left(p\right)\right)$ for $p\in\left[0,1\right]$.  Then, in Table \ref{tab:TsEqQIQIV}, we give the equilibria in each region that the line segment traverses.
The equilibrium
numbers refer to the derivations in the Appendix \ref{ssec:poolEq}.

\begin{table}[H]
\begin{centering}
\protect\protect\caption{Signaling game equilibria by region for 
a game that traverses between Quadrant I and Quadrant IV. Some of the equilibria
are feasible only for constrained beliefs $\mu$, specified in Appendix \ref{ssec:poolEq}. We argue that the equilibria in each region
marked by ({*}) will be selected.\label{tab:TsEqQIQIV}}

\par\end{centering}

\centering{}%
\begin{tabular}{|c|c|}
\hline 
Region  & Equilibria\tabularnewline
\hline 
\hline 
Quadrant I  & $\begin{array}{c}
\text{Equilibrium 3: Pool on }m_{L};\,\mu\text{ constrained;}\,\mathcal{R}\text{ plays }a_{T}\\
\text{*Equilibrium 8: Pool on }m_{H};\,\mu\text{ unconstrained;}\,\mathcal{R}\text{ plays }a_{T}
\end{array}$\tabularnewline
\hline 
$TB_H\left(p\right)=0$ Axis  & $\begin{array}{c}
\text{*Equilibrium 3: Pool on }m_{L};\,\mu\text{ constrained;}\,\mathcal{R}\text{ plays }a_{T}\\
\text{Equilibrium 8: Pool on }m_{H};\,\mu\text{ unconstrained;}\,\mathcal{R}\text{ plays }a_{T}\\
\text{Equilibrium 6: Pool on }m_{H};\,\mu\text{ constrained;}\,\mathcal{R}\text{ plays }a_{N}
\end{array}$\tabularnewline
\hline 
Quadrant IV  & $\begin{array}{c}
\text{*Equilibrium 3: Pool on }m_{L};\,\mu\text{ constrained;}\,\mathcal{R}\text{ plays }a_{T}\\
\text{Equilibrium 6: Pool on }m_{H};\,\mu\text{ constrained;}\,\mathcal{R}\text{ plays }a_{N}
\end{array}$\tabularnewline
\hline 
\end{tabular}
\end{table}

\begin{figure}[H]
\begin{centering}
\includegraphics[width=0.6\columnwidth]{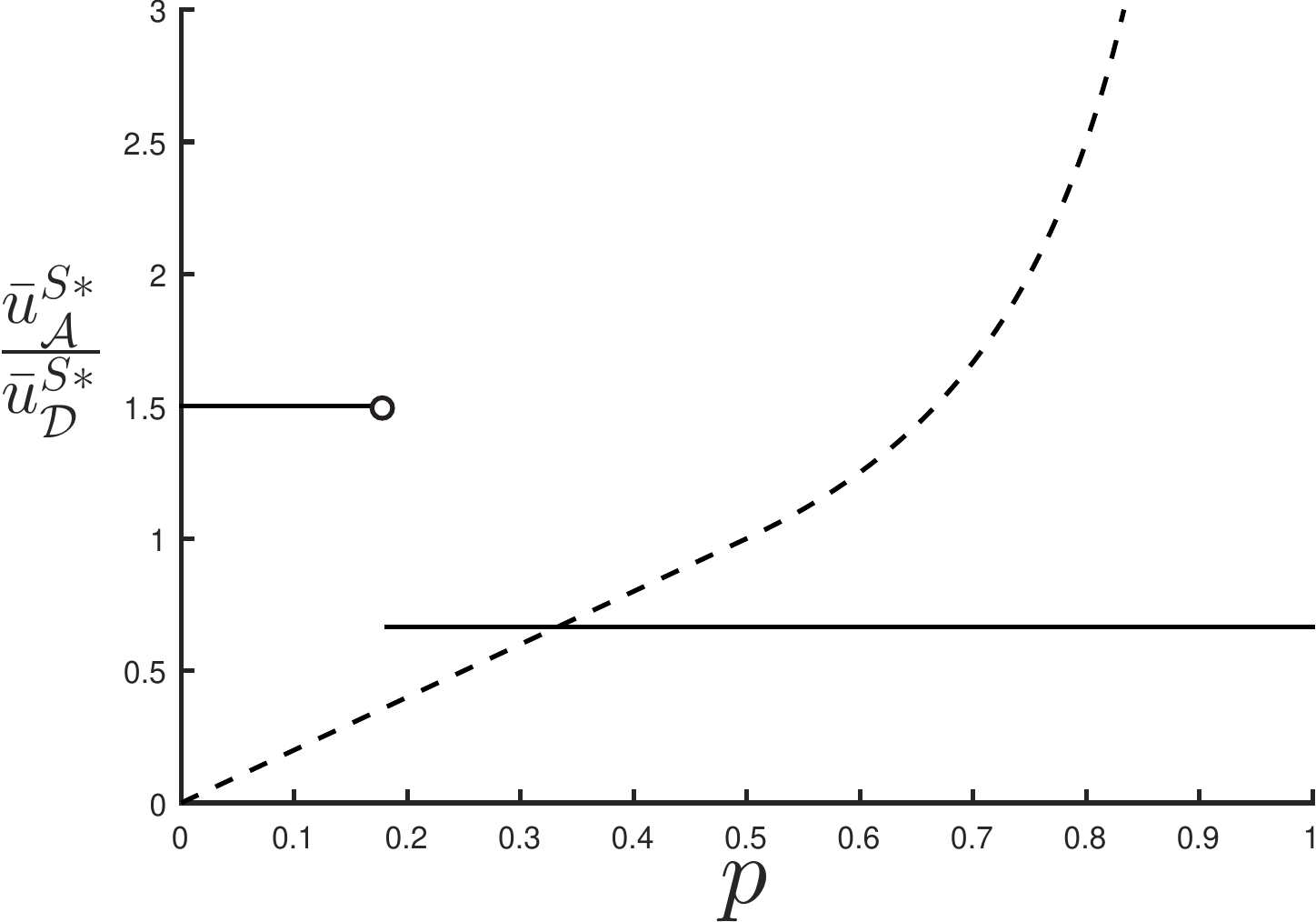} 
\par\end{centering}

\protect\protect\caption{$T^F$ and $T^S$ are combined on a single set of axis. In $T^S$ (the solid line), the independent variable is on the horizontal axis.  In $T^F$ (the dashed line), the independent variable is on the vertical axis.  The intersection of the two curves represents the Gestalt equilibrium.\label{fig:TfAndTs}}
\end{figure}

If $p$ is such that the signaling game is played in Quadrant I, then both senders prefer pooling on $m_{H}$. By the \emph{first mover
advantage}, they will select Equilibrium
8. On the border between Quadrant I and Quadrant IV, $\mathcal{A}$
and $\mathcal{D}$ both prefer an equilibrium in which $\mathcal{R}$
plays $a_{T}$. If they pool on $m_{L}$, this is guaranteed. If they
pool on $m_{H}$, however, $\mathcal{R}$ receives equal utility for
playing $a_{T}$ and $a_{N}$; thus, the senders cannot guarantee
that the receiver will play $a_{T}$. Here, we assume that the senders
maximize their worst-case utility, and thus pool on $m_{L}$. This
is Equilibrium 3. Finally, in Quadrant IV, both senders prefer to
be trusted, and so select Equilibrium 3. From the table, we can see
that the utilities will have a jump at the border between Quadrant I and Quadrant IV.
The solid line in Fig. \ref{fig:TfAndTs} plots the ratio $\bar{u}_{\mathcal{A}}^{S*}/\bar{u}_{\mathcal{D}}^{S*}$ of the utilities as a function of $p$.

Next, consider the mapping $p=T^{F}\left(\bar{u}_{\mathcal{D}}^{S*},\bar{u}_{\mathcal{A}}^{S*}\right)$.
%The combination of Equations (\ref{eq:p}) and (\ref{eq:FlipNE1})-(\ref{eq:FlipNE4}) specify%that the output $p$ is given by%\begin{equation}%p=\begin{cases}%0,\text{ if } & \bar{u}_{\mathcal{A}}^{S*}\leq0\\%\frac{\alpha_{\mathcal{D}}\bar{u}_{\mathcal{A}}^{S*}}{2\alpha_{\mathcal{A}}\bar{u}_{\mathcal{D}}^{S*}},\text{ if } & \frac{\alpha_{\mathcal{A}}}{\bar{u}_{\mathcal{A}}^{S*}}\geq\frac{\alpha_{\mathcal{D}}}{\bar{u}_{\mathcal{D}}^{S*}}\text{ and }\bar{u}_{\mathcal{A}}^{S*},\bar{u}_{\mathcal{D}}^{S*}>0\\%1-\frac{\alpha_{\mathcal{A}}\bar{u}_{\mathcal{D}}^{S*}}{2\alpha_{\mathcal{D}}\bar{u}_{\mathcal{A}}^{S*}},\text{ if } & \frac{\alpha_{\mathcal{A}}}{\bar{u}_{\mathcal{A}}^{S*}}\leq\frac{\alpha_{\mathcal{D}}}{\bar{u}_{\mathcal{D}}^{S*}}\text{ and }\bar{u}_{\mathcal{A}}^{S*},\bar{u}_{\mathcal{D}}^{S*}>0\\%1,\text{ if } & \bar{u}_{\mathcal{A}}^{S*}>0\text{ and }\bar{u}_{\mathcal{D}}^{S*}\leq0%\end{cases}.\label{eq:TsCases}%\end{equation}As
As we have noted, $p$ depends only on the ratio $\bar{u}_{\mathcal{A}}^{S*}/\bar{u}_{\mathcal{D}}^{S*}$%
\footnote{When $\bar{u}_{\mathcal{A}}^{S*}=\bar{u}_{\mathcal{D}}^{S*}=0$, we
define that ratio to be equal to zero, since this will yield $f_{\mathcal{A}}=0$
and $p=0$, as in Equations (\ref{eq:expUtilAttNo}) and (\ref{eq:p}). When 
$\bar{u}_{\mathcal{D}}^{S*}=0$ and $\bar{u}_{\mathcal{A}}^{S*}>0$, it is convenient to consider the ratio to be positively infinite since this is consistent with $p\to1$.} . Indeed, it is continuous in that ratio when the outcome at the
endpoints is appropriately defined. %are defined appropriately. %Fig. \ref{fig:TfPlot}, in which we put the independent variable on the vertical axis and the dependent variable on the horizontal axis, shows the dependence of $p$ on $\bar{u}_{\mathcal{A}}^{S*}/\bar{u}_{\mathcal{D}}^{S*}$.
This mapping
is represented by the dashed line in Fig. \ref{fig:TfAndTs}, with the
independent variable on the vertical axis. 

We seek a fixed-point, in which $p=T^{F}\left(\bar{u}_{\mathcal{D}}^{S*},\bar{u}_{\mathcal{A}}^{S*}\right)$
and $\left(\bar{u}_{\mathcal{D}}^{S*},\bar{u}_{\mathcal{A}}^{S*}\right)=T^{S}\left(p\right)$.
This %of course, is the intersection of the curves in Fig. \ref{fig:TsEqQIQIVRatio} and Fig. \ref{fig:TfPlot}, which is
shown by the intersection of the solid and dashed curves plotted in Fig.
\ref{fig:TfAndTs}. At these points, the mappings between the signaling and the \texttt{FlipIt}
games are mutually satisfied, and we have a Gestalt equilibrium.\footnote{Note that this example featured a 
discontinuity in signaling game utilities on the border between equilibrium regions.
Interestingly, even when the pooling equilibria differ between regions, it is possible
that the equilibrium on the border admits a mixed strategy that provides continuity
between the different equilibria in the two regions, and thus makes $T^S$
continuous. This could allow $\mathbf{G_{CC}}$ to have multiple Gestalt 
equilibria.}

\section{Cloud Control Application}
\label{sec:Simu}
%In Section \ref{sec:Intro}, we discussed how \texttt{CloudControl} games are poised to have a large importance due to the benefit that can be offered by cyber-physical systems. 

%% This paragraph after initial submission:[Removed 8/25]
%Thus far, we have described the cyber layer of a cloud-enabled CPS using
%a \texttt{FlipIt} game and the interaction of the cyber and physical
%layers using a signaling game.  But what is the role of the physical layer?
%The physical layer determines the costs and benefits of controling the device.
%In the language of optimal control, these costs or benefits could be 
%modeled using a functional such as the linear combination of penalties on
%state and control that are considered in a linear-quadratic regulator (LQR).
%Because different physical devices will be governed by different dynamic
%systems and different cost structures, they will generate different utilities
%for controlling the device. In this section, we consider a particular physical
%device: an automated vehicle equipped to access the cloud.

In this section, we describe one possible application of our model: a cyber-physical system composed of autonomous 
vehicles with some on-board control but also with the ability to 
trust commands from the cloud.
Access to the cloud can offer automated vehicles several
benefits \cite{kehoe2015survey}. First, it allows access to massive
computational resources - \emph{i.e.,} \emph{infrastructure as a service}
(\emph{IaaS}). (See \cite{bhardwaj2010cloud}.) Second, it allows
access to large datasets. These datasets can offer super-additive benefits
to the sensing capabilities of the vehicle itself, as in the case
of the detailed road and terrain maps that automated cars such as
those created by Google and Delphi combine with data collected by
lidar, radar and vision-based cameras \cite{delphi,guizzo2011google}.
Third, interfacing with the cloud allows access to data collected
or processed by humans through crowd-sourcing applications; consider,
for instance, location-based services \cite{sampigethaya2005caravan,sampigethaya2007amoeba}
that feature recommendations from other users. Finally, the cloud
can allow vehicles to collectively learn through experience \cite{kehoe2015survey}.
%As we have seen, however, utilizing the cloud comes with security concerns.

Attackers may attempt to influence cloud control of the vehicle through
several means. In one type of attack, adversaries may be able to \emph{steal
or infer cryptographic keys} that allow them authorization into the
network. These attacks are of the \emph{complete compromise }and \emph{stealth}
types that are studied in the \texttt{FlipIt }framework \cite{vanDijk2013Flip}, \cite{bowers2012defending}
and thus are appropriate for a \texttt{CloudControl} game. \texttt{FlipIt
}also provides the ability to model \emph{zero-day exploit}s, vulnerabilities
for which a patch is not currently available. 
%Finally, we can consider attacks upon a cloud that controls unmanned vehicles through \emph{malware injections}: insertions of malicious virtual machines with the purpose of eavesdropping, functionally altering or blocking the function of a device \cite{jensen2009technical}. 
Each of these types of
attacks on the cloud pose threats to unmanned vehicle security and
involve the complete compromise and steathiness that motivate the
\texttt{FlipIt }framework. 

\subsection{Dynamic Model for Cloud Controlled Unmanned Vehicles}

%In Section \ref{sec:SysMod}, we described our \texttt{CloudControl} model for $\mathbf{G_{CC}}$. This model combined a \texttt{FlipIt} game for control of the cloud with a signaling game between the cloud controller and a device. We also specified that the model of the physical component would depend on the application. 
In this subsection, we use a dynamic
model of an autonomous car to illustrate one specific context in which
a cloud-connected device could be making a decision of whether to 
trust the commands that it would receive or to follow its own 
on-board control.

\begin{figure}[h]
\begin{center}
\includegraphics[width=0.5\columnwidth]{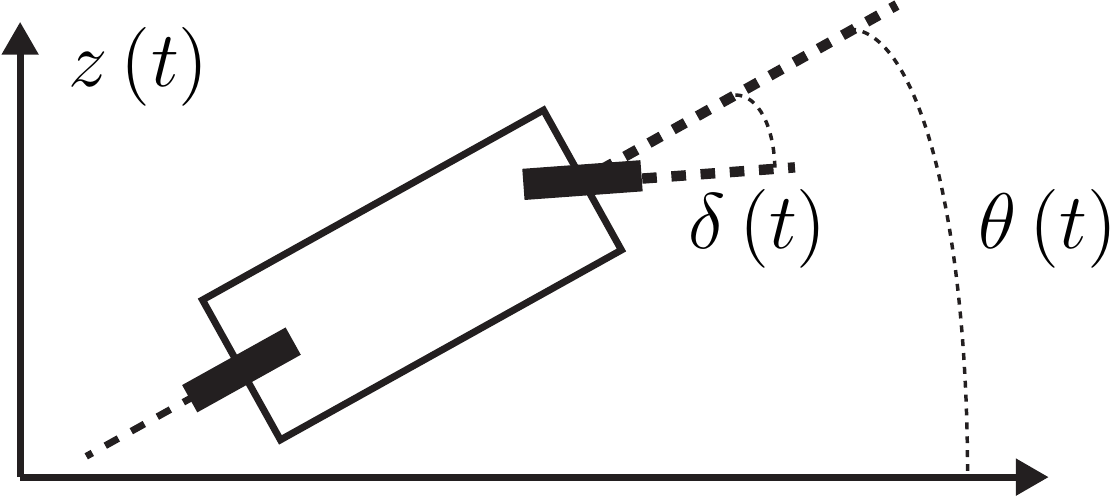}

\caption{A bicycle model is a type of representation of vehicle steering control.  Here, $\delta\left(t\right)$ is used to denote the angle between the orientation of the front wheel and the heading $\theta\left(t\right)$.  The deviation of the vehicle from a straight line is given by $z\left(t\right)$}
\label{fig:CaseStud1}
\end{center}
\end{figure}

We consider a car moving in two-dimensional space with a fixed speed
$v_0$ but with steering that can be controlled. (See Fig. \ref{fig:CaseStud1}, which illustrates the ``bicycle model'' of steering control
from \cite{astrom2010feedback}.) For simplicity, assume that we are
interested in the car's deviation from a straight line. (This line
might, \emph{e.g.}, run along the center of the proper driving lane.)
Let $z\left(t\right)$ denote the car's vertical distance from the
horizontal line, and let $\theta\left(t\right)$ denote the heading
of the car at time $t$. The state of the car can be represented by
a two-dimensional vector $w\left(t\right)\triangleq\left[\begin{array}{cc}
z\left(t\right) & \theta\left(t\right)\end{array}\right]^{T}$. Let $\delta\left(t\right)$ denote the angle between the orientation
of the front wheel - which implements steering - and the orientation
of the length of the car. We can consider $\delta\left(t\right)$
to be the input to the system. Finally, let $y\left(t\right)$ represent
a vector of outputs available to the car's control system. The self-driving
cars of both Google and Delphi employ radar, lidar, and vision-based
cameras for localization. Assume that these allow accurate measurement
of both states, such that $y_{1}\left(t\right)=z\left(t\right)$ and
$y_{2}\left(t\right)=\theta\left(t\right)$. If the car stays near
$w\left(t\right)=\left[\begin{array}{cc}
0 & 0\end{array}\right]^{T}$, then we can approximate the system with a linear model. Let $a$
and $b$ denote the distances from the rear wheel to the center of
gravity and the rear wheel to the front wheel of the car, respectively. Then the linearized system is given in \cite{astrom2010feedback}
by the equations:
\begin{equation}
\frac{d}{dt}\left[\begin{array}{c}
z\left(t\right)\\
\theta\left(t\right)
\end{array}\right]=\left[\begin{array}{cc}
0 & v_{0}\\
0 & 0
\end{array}\right]\left[\begin{array}{c}
z\left(t\right)\\
\theta\left(t\right)
\end{array}\right]+\left[\begin{array}{c}
\frac{av_{0}}{b}\\
\frac{v_{0}}{b}
\end{array}\right]\delta\left(t\right),
\end{equation}

\begin{equation}
\left[\begin{array}{c}
y_{1}\left(t\right)\\
y_{2}\left(t\right)
\end{array}\right]=\left[\begin{array}{cc}
1 & 0\\
0 & 1
\end{array}\right]\left[\begin{array}{c}
z\left(t\right)\\
\theta\left(t\right)
\end{array}\right].
\end{equation}

\subsection{Control of Unmanned Vehicle}

Assume that the unmanned car has some capacity for automatic control without
the help of the cloud, but that the cloud typically provides more
advanced navigation. 

Specifically, consider a control system onboard the unmanned vehicle
designed to return it to the equilibrium $w\left(t\right)=\left[\begin{array}{cc}
0 & 0\end{array}\right]^{T}$. Because the car has access to both of the states, it can implement
a state-feedback control. Consider a linear, time-invariant control
of the form
\begin{equation}
\delta_{car}\left(t\right)=-\left[\begin{array}{cc}
k_{1} & k_{2}\end{array}\right]\left[\begin{array}{c}
z\left(t\right)\\
\theta\left(t\right)
\end{array}\right].\label{eq:feedback}
\end{equation}
This proportional
control results in the closed-loop system
\begin{equation}
\frac{d}{dt}\left[\begin{array}{c}
z\left(t\right)\\
\theta\left(t\right)
\end{array}\right]=\left(\left[\begin{array}{cc}
0 & v_{0}\\
0 & 0
\end{array}\right]-\left[\begin{array}{c}
\frac{av_{0}}{b}\\
\frac{v_{0}}{b}
\end{array}\right]\left[\begin{array}{cc}
k_{1} & k_{2}\end{array}\right]\right)\left[\begin{array}{c}
z\left(t\right)\\
\theta\left(t\right)
\end{array}\right].\label{eq:carControl}
\end{equation}

The unmanned car $\mathcal{R}$ may also elect to obtain data or computational
resources from the cloud. Typically, this additional access would improve the control
of the car. The cloud administrator (defender $\mathcal{D}$), however,
may issue faulty commands or there may be a breakdown in communication
of the desired signals. In addition, the cloud may be compromised
by $\mathcal{A}$ in a way that is stealthy.
Because of these factors, $\mathcal{R}$ sometimes benefits from rejecting
the cloud's command and relying on its own navigational abilities.
Denote the command issued by the cloud at time $t$ by 
$\delta_{cloud}\left(t\right)
\in{
\delta_{\mathcal{A}}\left(t\right),
\delta_{\mathcal{D}}\left(t\right)
}$, depending on who controls the cloud.
With this command, the system is given by
\begin{equation}
\frac{d}{dt}\left[\begin{array}{c}
z\left(t\right)\\
\theta\left(t\right)
\end{array}\right]=\left[\begin{array}{cc}
0 & v_{0}\\
0 & 0
\end{array}\right]\left[\begin{array}{c}
z\left(t\right)\\
\theta\left(t\right)
\end{array}\right]+\left[\begin{array}{c}
\frac{av_{0}}{b}\\
\frac{v_{0}}{b}
\end{array}\right]\delta_{cloud}\left(t\right).\label{eq:cloudControl}
\end{equation}

\subsection{Filter for High Risk Cloud Commands}

%In Section \ref{sec:SysMod}, we specified that messages could be filtered by a device as low or high risk. 
In cloud control of an unmanned
vehicle, the self-navigation state feedback input given by $\delta_{car}\left(t\right)$
in Equation (\ref{eq:feedback}) represents the control that is expected
by the vehicle given its state. If the signal from the cloud differs
significantly from the signal given by the self-navigation system,
then the vehicle may classify the message as ``high-risk.'' Specifically,
define a \emph{difference threshold} $\tau$, and let
\begin{equation}
m=\begin{cases}
m_{H},\text{ if } & \left|\delta_{cloud}\left(t\right)-\delta_{car}\left(t\right)\right|>\tau\\
m_{L},\text{ if } & \left|\delta_{cloud}\left(t\right)-\delta_{car}\left(t\right)\right|\leq\tau
\end{cases}.\label{eq:filter}
\end{equation}
Equation (\ref{eq:filter}) translates the actual command from the
cloud (controlled by $\mathcal{D}$ or $\mathcal{A}$) into a message
in the cloud signaling game. 

%\subsection{Control Model for Partially-Trusted Cloud}

\begin{figure}[h]
\begin{center}
\includegraphics[scale=0.6]{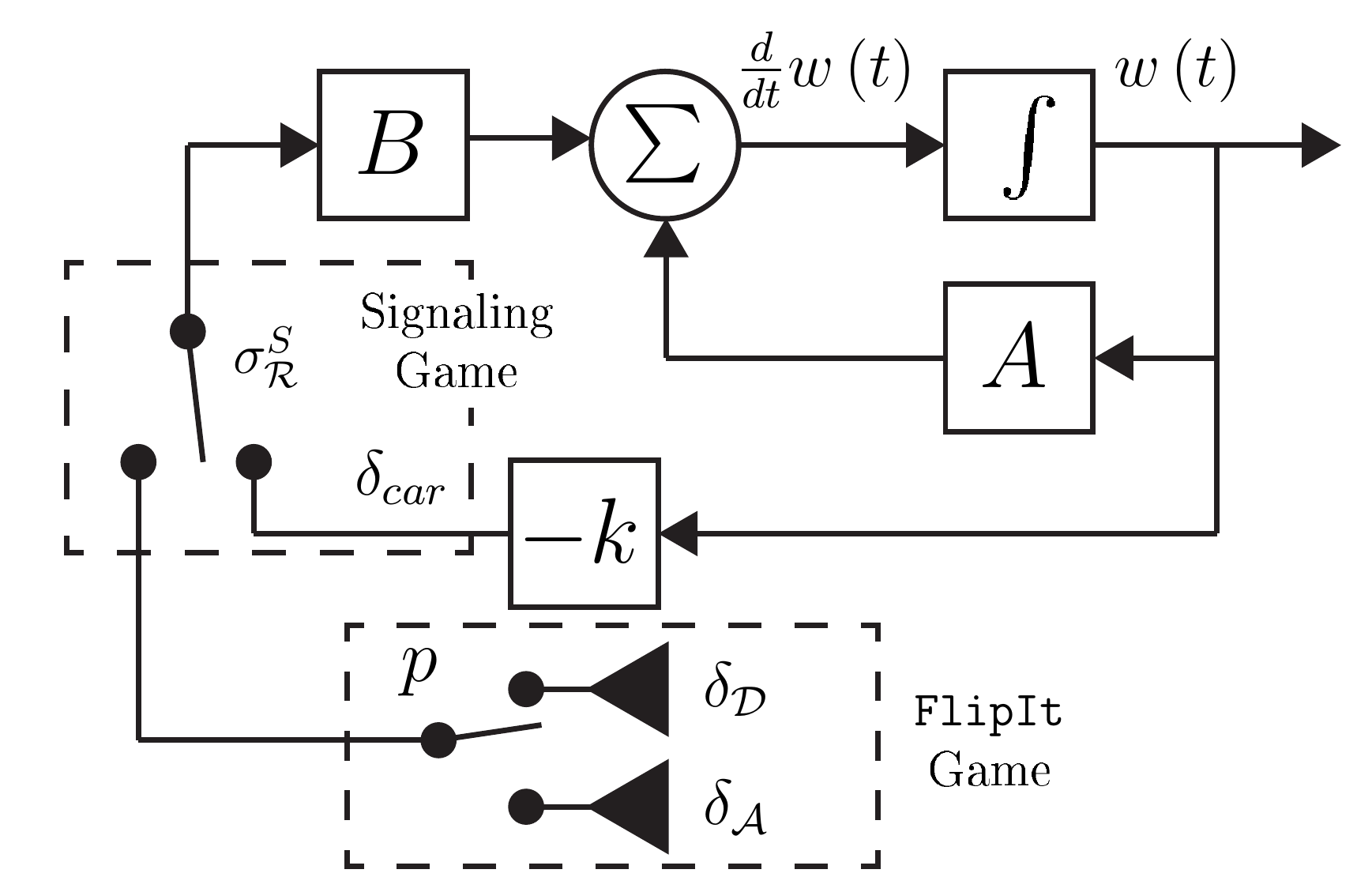}

\caption{Block-diagram model for unmanned vehicle navigation control.  
At any time, the vehicle uses strategy $\sigma^S_\mathcal{R}$ to decide whether to follow its own control or the control
signal from the cloud, which may be $\delta_\mathcal{A}$ or $\delta_\mathcal{D}$, depending on the probabilities $p$, $1-p$ with which $\mathcal{A}$ and $\mathcal{D}$ control the cloud. Its own control signal, $\delta_{car}$, is obtained via feedback control.} 
\label{fig:CaseStud2}
\end{center}
\end{figure}

Equations (\ref{eq:carControl}) and (\ref{eq:cloudControl}) give the
dynamics of the unmanned car electing to trust and not trust
the cloud. Based on these equations, Fig. \ref{fig:CaseStud2} illustrates the
combined self-navigating and cloud controlled system for vehicle control. 

%\subsection{Results for Cloud Control of Unmanned Vehicle}
%
%{[}THIS SECTION FOR SATURDAY{]}
%%%%%%%%%%%%%%%%%%%%%%%%%%%%%%%%%%%%%% 
\section{Conclusion and Future Work}
\label{sec:Conclu}
In this paper, we have proposed a general framework for the interaction between an attacker, cloud administrator/defender, and cloud-connected device.  We have described the struggle for control of the cloud using the \texttt{FlipIt} game and the interaction between the cloud and the connected device using a traditional signaling game.  Because these two games are played by prior commitment, they are coupled.  We have defined a new equilibrium concept - \emph{i.e., Gestalt equilibrium}, which defines a solution to the combined game using a fixed-point equation.  After illustrating various parameter regions under which the game may be played, we solved the game in a sample parameter region.  Finally, we showed how the framework may be applied to unmanned vehicle control.

Several directions remain open for future work.  First, the physical component of the cyber-physical system can be further examined.  Tools from optimal control such as the linear-quadratic regulator could offer a rigerous framework for defining the costs associated with the physical dynamic system, which in turn would define the payoffs of the signaling game.  Second, future work could search for conditions under which a Gestalt equilibrium of the \texttt{CloudControl} game is guaranteed to exist.  Finally, devices that use this framework should be equipped to learn online.  Towards that end, a learning algorithm could be developed that is guaranteed to converge to the Gestalt equilibrium. Together with the framework developed in the present paper, these directions would help to advance our ability to secure cloud-connected and cyber-physical systems.
%%%%%%%%%%%%%%%%%%%%%%%%%%%%%%%%%%%%%%%%%%%%%%%%%%%%%%%%%%%
\bibliographystyle{amsplain}
\bibliography{gs1}
%%%%%%%%%%%%%%%%%%%%%%%%%%%%%%%%%%%%%%%%%%%%%%%%%%%%%%%%%%%
\appendix
%%%%%%%%%%%%%%%%%%%%%%%%%%%%%%%%%%%%%% 
\section{Derivation of Signaling Game Equilibria} 
\label{apdx}
In this appendix, we solve for the equilibria of $\mathbf{G_{S}}$.

\subsection{Separating Equilibria}

\label{ssec:sepEq}

First, we search for separating equilibria of $\mathbf{G_{S}}$. In
separating equilibria, $\mathcal{R}$ knows with certainty the type
of the cloud.

\subsubsection{$\mathcal{D}$ plays $m_{L}$ and $\mathcal{A}$ plays $m_{H}$}

If $\mathcal{D}$ plays $m_{L}$ (as a pure strategy) and $\mathcal{A}$
plays $m_{H}$, then the receiver rejects any $m_{H}$ according to
assumption A2. The best action for $\mathcal{A}$ is to deviate to
$m_{L}$. Thus, this is not an equilibrium.

\subsubsection{$\mathcal{D}$ plays $m_{H}$ and $\mathcal{A}$ plays $m_{L}$}

If $\mathcal{D}$ plays $m_{H}$ and $\mathcal{A}$ plays $m_{L}$,
the $\mathcal{R}$'s best response depends on the utility parameters.
If $u_{\mathcal{R}}^{S}\left(\theta_{\mathcal{A}},m_{L},a_{T}\right)\leq u_{\mathcal{R}}^{S}\left(\theta_{\mathcal{A}},m_{L},a_{N}\right)$
and $u_{\mathcal{R}}^{S}\left(\theta_{\mathcal{D}},m_{H},a_{T}\right)\leq u_{\mathcal{R}}^{S}\left(\theta_{\mathcal{D}},m_{H},a_{N}\right)$,
then $\mathcal{R}$ plays $a_{N}$ in response to both messages. There
is no incentive to deviate. Denote this separating equilibrium as
\emph{Equilibrium \#2}.

If $u_{\mathcal{R}}^{S}\left(\theta_{\mathcal{A}},m_{L},a_{T}\right)\leq u_{\mathcal{R}}^{S}\left(\theta_{\mathcal{A}},m_{L},a_{N}\right)$
and $u_{\mathcal{R}}^{S}\left(\theta_{\mathcal{D}},m_{H},a_{T}\right)>u_{\mathcal{R}}^{S}\left(\theta_{\mathcal{D}},m_{H},a_{N}\right)$,
then $a_{N}$ is within the set of best responses to $m_{L}$, whereas
$a_{T}$ is the unique best response to $m_{H}$. Assuming that he
prefers to certainty receive a higher utility, $\mathcal{A}$ deviates
to $m_{H}$.

If $u_{\mathcal{R}}^{S}\left(\theta_{\mathcal{A}},m_{L},a_{T}\right)>u_{\mathcal{R}}^{S}\left(\theta_{\mathcal{A}},m_{L},a_{N}\right)$
and $u_{\mathcal{R}}^{S}\left(\theta_{\mathcal{D}},m_{H},a_{T}\right)\leq u_{\mathcal{R}}^{S}\left(\theta_{\mathcal{D}},m_{H},a_{N}\right)$,
then $a_{N}$ is within the set of best responses to $m_{H}$, whereas
$a_{T}$ is the unique best response to $m_{L}$. Thus, $\mathcal{D}$
deviates to $m_{L}$.

If $u_{\mathcal{R}}^{S}\left(\theta_{\mathcal{A}},m_{L},a_{T}\right)>u_{\mathcal{R}}^{S}\left(\theta_{\mathcal{A}},m_{L},a_{N}\right)$
and $u_{\mathcal{R}}^{S}\left(\theta_{\mathcal{D}},m_{H},a_{T}\right)>u_{\mathcal{R}}^{S}\left(\theta_{\mathcal{D}},m_{H},a_{N}\right)$,
then $\mathcal{R}$ plays $a_{T}$ in response to both messages. We
have assumed, however, that $\mathcal{A}$ prefers to be trusted on
$m_{H}$ compared to being trusted on $m_{L}$ (A4), so $\mathcal{A}$
deviates and this is not an equilibrium.

\subsection{Pooling Equilibria}

\label{ssec:poolEq}

Next, we search for pooling equilibria of $\mathbf{G_{S}}$. In pooling
equilibria, $\mathcal{R}$ relies only on the prior probabilities
$p$ and $1-p$ in order to form his belief about the type of the
cloud. The existence of pooling equilibria depend essentially on the
trust benefits $TB_{H}\left(p\right)$ and $TB_{L}\left(p\right).$

\subsubsection{Pooling on $m_{L}$}

If $TB_{L}\left(p\right)<0$, then $\mathcal{R}$'s best response
is $a_{N}$. This will only be an equilibrium if his best response
to $m_{H}$ would also be $a_{N}$. This is the case only when the
belief satisfies 

\begin{equation}
\begin{array}{c}
\mu\left(\theta_{\mathcal{A}}\,|\, m_{H}\right)u_{\mathcal{R}}\left(\theta_{\mathcal{A}},m_{H},a_{T}\right)+\left(1-\mu\left(\theta_{\mathcal{A}}\,|\, m_{H}\right)\right)u_{\mathcal{R}}\left(\theta_{\mathcal{D}},m_{H},a_{T}\right)\\
\leq\mu\left(\theta_{\mathcal{A}}\,|\, m_{H}\right)u_{\mathcal{R}}\left(\theta_{\mathcal{A}},m_{H},a_{N}\right)+\left(1-\mu\left(\theta_{\mathcal{A}}\,|\, m_{H}\right)\right)u_{\mathcal{R}}\left(\theta_{\mathcal{D}},m_{H},a_{N}\right)
\end{array}.\label{eq:beliefLow}
\end{equation}
Moreover, this can only be an equilibrium when neither $\mathcal{A}$
nor $\mathcal{D}$ have an incentive to deviate: \emph{i.e.}, when
\begin{equation}
u_{\mathcal{A}}^{S}\left(m_{H},a_{N}\right)\leq u_{\mathcal{A}}^{S}\left(m_{L},a_{N}\right)\;\text{and\;}u_{\mathcal{D}}^{S}\left(m_{H},a_{N}\right)\leq u_{\mathcal{D}}^{S}\left(m_{L},a_{N}\right).
\end{equation}
If these conditions are satisfied, then denote this equilibrium by
\emph{Equilibrium \#1}.

If $TB_{L}\left(p\right)\geq0$, then $\mathcal{R}$'s best response
us $a_{T}$. Whether this represents an equilibrium depends on if
$\mathcal{A}$ or $\mathcal{D}$ have incentives to deviate from $m_{L}$.
If $u_{\mathcal{D}}^{S}\left(m_{H},a_{T}\right)\leq u_{\mathcal{D}}^{S}\left(m_{L},a_{T}\right)$
and $u_{\mathcal{A}}^{S}\left(m_{H},a_{T}\right)\leq u_{\mathcal{A}}^{S}\left(m_{L},a_{T}\right)$,
then neither has an incentive to deviate. This is \emph{Equilibrium
\#5}. If one of these inequalities does \emph{not} hold, then the
player who prefers $m_{H}$ to $m_{L}$ will deviate if $\mathcal{R}$
would play $a_{T}$ in response to the deviation. The equilibrium
condition is narrowed to when the belief makes $\mathcal{R}$ not
trust $m_{H}$; when Equation (\ref{eq:beliefLow}) is satisfied.
Call this \emph{Equilibrium \#3}.

\subsubsection{Pooling on $m_{H}$}

The pattern of equilibria for pooling on $m_{H}$ follows a similar
structure to the pattern of equilibria for pooling on $m_{L}$.

If $TB_{H}\left(p\right)<0$, then $\mathcal{R}$'s best response
is $a_{N}$. This will only be an equilibrium if his best response
to $m_{L}$ would also be $a_{N}$. This is the case only when the
belief satisfies 
\begin{equation}
\begin{array}{c}
\mu\left(\theta_{\mathcal{A}}\,|\, m_{L}\right)u_{\mathcal{R}}\left(\theta_{\mathcal{A}},m_{L},a_{T}\right)+\left(1-\mu\left(\theta_{\mathcal{A}}\,|\, m_{L}\right)\right)u_{\mathcal{R}}\left(\theta_{\mathcal{D}},m_{L},a_{T}\right)\\
\leq\mu\left(\theta_{\mathcal{A}}\,|\, m_{L}\right)u_{\mathcal{R}}\left(\theta_{\mathcal{A}},m_{L},a_{N}\right)+\left(1-\mu\left(\theta_{\mathcal{A}}\,|\, m_{L}\right)\right)u_{\mathcal{R}}\left(\theta_{\mathcal{D}},m_{L},a_{N}\right)
\end{array}.\label{eq:beliefHigh}
\end{equation}
To guarantee that $\mathcal{A}$ and $\mathcal{D}$ do not deviate,
we require 

\begin{equation}
u_{\mathcal{A}}^{S}\left(m_{H},a_{N}\right)\geq u_{\mathcal{A}}^{S}\left(m_{L},a_{N}\right)\;\text{and\;}u_{\mathcal{D}}^{S}\left(m_{H},a_{N}\right)\geq u_{\mathcal{D}}^{S}\left(m_{L},a_{N}\right).
\end{equation}
If these conditions are satisfied, then we have \emph{Equilibrium
\#6.}

If $TB_{H}\geq0$, then $\mathcal{R}$'s best response is $a_{T}$.
If $u_{\mathcal{D}}^{S}\left(m_{H},a_{T}\right)\geq u_{\mathcal{D}}^{S}\left(m_{L},a_{T}\right)$
and $u_{\mathcal{A}}^{S}\left(m_{H},a_{T}\right)\geq u_{\mathcal{A}}^{S}\left(m_{L},a_{T}\right)$,
then neither $\mathcal{A}$ nor $\mathcal{D}$ have an incentive to
deviate. Call this \emph{Equilibrium \#8}. If one of these inequalities
does not hold, then the belief must satisfy Equation (\ref{eq:beliefHigh})
for an equilibrium to be sustained. Denote this equilibrium by \emph{Equilibrium
\#7}.

%%%%%%%%%%%%%%%%%%%%%%%%%%%%%%%%%%%%%% 
%\section{Separating Equilibria for which }
%\label{SepDHAL}
%\input{SepDHAL}
%%%%%%%%%%%%%%%%%%%%%%%%%%%%%%%%%%%%%% 
%\section{Pooling Equilibria}
%$u_{\mathcal{D}}^{S}\left(m_{H}\right)=0$,
%$u_{\mathcal{A}}^{S}\left(m_{H}\right)=0$}
%\label{PoolL}
%\input{PoolL}
%%%%%%%%%%%%%%%%%%%%%%%%%%%%%%%%%%%%%% 
%\section{Pooling Equilibria for which $u_{\mathcal{D}}^{S}\left(m_{H}\right)=1$,
%$u_{\mathcal{A}}^{S}\left(m_{H}\right)=1$}
%\label{PoolH}
%\input{PoolH}
%%%%%%%%%%%%%%%%%%%%%%%%%%%%%%%%%%%%%%%%%%

\end{document}